\documentclass[twocolumn,showpacs,pra,aps,superscriptaddress,floatfix]{revtex4-1}
\usepackage{graphicx}
\usepackage{braket} 
\usepackage{float} 
\usepackage{natbib}
\usepackage[latin1]{inputenc}           
\usepackage{bm} 
\usepackage{amsmath}
\usepackage{amssymb}
\usepackage{mathtools}
\usepackage{bbold}
\usepackage{comment}
\usepackage{dsfont}

\usepackage{color}
 \usepackage{soul}

 \definecolor{darkred}{rgb}{0.8,0.1,0.1}
 \definecolor{DARKRED}{rgb}{0.8,0.1,0.1}
 \definecolor{darkblue}{rgb}{0.1,0.1,0.7}
 \definecolor{bleudefrance}{rgb}{0.19, 0.55, 0.91}
 
\usepackage{colordvi}

\definecolor{orange}{RGB}{204,102,0}

\definecolor{redjesus}{RGB}{204,0,102}

\definecolor{DodgerBlue}{RGB}{0, 90, 156}

\definecolor{greenfernando}{RGB}{51, 102, 0}
\newcommand{\codi}[1]{\textcolor{violetdiego}{#1}}
\definecolor{violetdiego}{RGB}{153,0, 153}
\usepackage{caption}
\usepackage{subcaption}

\begin{document}

\title{Uhlmann phase in composite systems with entanglement}

\author{J. Villavicencio}
 \affiliation{Centro de Nanociencias y Nanotecnolog\'ia, Universidad
 Nacional Aut\'onoma de M\'exico, Apartado Postal 14, 22800 Ensenada, B.C., M\'exico}
 \affiliation{Facultad de Ciencias, Universidad Aut\'onoma de Baja
 California, 22800 Ensenada, B.C., M\'exico}

 \author{E. Cota}
 \affiliation{Centro de Nanociencias y Nanotecnolog\'ia, Universidad
 Nacional Aut\'onoma de M\'exico, Apartado Postal 14, 22800 Ensenada, B.C., M\'exico}
 \author{F. Rojas}
 \affiliation{Centro de Nanociencias y Nanotecnolog\'ia, Universidad
 Nacional Aut\'onoma de M\'exico, Apartado Postal 14, 22800 Ensenada, B.C., M\'exico}

 \author{J.~A. Maytorena}
 \affiliation{Centro de Nanociencias y Nanotecnolog\'ia, Universidad
 Nacional Aut\'onoma de M\'exico, Apartado Postal 14, 22800 Ensenada, B.C., M\'exico}

 \author{D. Morachis Galindo}
 \affiliation{Centro de Nanociencias y Nanotecnolog\'ia, Universidad
 Nacional Aut\'onoma de M\'exico, Apartado Postal 14, 22800 Ensenada, B.C., M\'exico}

\date{\today}
\begin{abstract}
%

We study the geometric Uhlmann phase of entangled mixed states in a composite system made of two coupled spin-$\frac 1 2$ particles with
a magnetic field acting on one of them. Within a depolarizing channel setup, an exact analytical expression for such a phase in each subsystem is derived.
We find an explicit connection to the concurrence of the depolarizing channel density matrix, which allows to characterize the features of the Uhlmann phase in terms of the degree of entanglement in the system. 
In the space of field direction and coupling parameter, it exhibits a phase singularity revealing a topological transition between orders with different winding numbers. The transition occurs for fields lying in the equator of the sphere of directions and at critical values of the coupling which can be controlled by tuning the depolarization strength.
Notably, under these conditions
the concurrence of the composite system is bounded to the range $[0,1/2]$. We also compare the calculated Uhlmann phase
to an interferometric phase, which has been formulated as an alternative for density matrices. The latter does not present a phase vortex, although they coincide in the weak entanglement regime, for vanishing depolarization (pure states). Otherwise they behave clearly different in the strong entanglement regime.

\end{abstract}
\maketitle

\section{Introduction}

Since the seminal work by Berry \cite{Berry84},  geometric phases in pure quantum systems have played an important role in condensed matter physics \cite{vanderbilt_2018}. 
When quantum states undergo  an adiabatic  cyclic evolution,  they acquire a nontrivial phase which depends on the path followed by the state vector in parameter space, in addition to the usual dynamical phase. 
These phases are robust under perturbations in the dynamics of the quantum state, due to their geometric nature.
The latter has become a desirable feature, for example, in the context of quantum information, since it allows to 
design fault-tolerant quantum logic gates applied to quantum computation \cite{jones2000}.
Nevertheless, the presence of entanglement in the components of computational setups, requires an understanding and control of geometric phases for mixed states. 
The Uhlmann phase \cite{uhlmannrmp86,uhlmannlmp91} provides an extension of the Berry phase from pure states to the realm of density matrices.
The relevance of the Uhlmann phase to describe the geometric phases in mixed systems has been stressed recently by Viyuela \textit{et al.} \cite{viyuelaprl14,Viyuela_2015}.
In the latter work it was shown that the Uhlmann phase exhibits a topological structure, at both zero and finite temperatures, 
for several paradigmatic one-dimensional fermionic systems: polyacetilene (SSH model) \cite{suprl79,shorty_2014}, Creutz ladder  \cite{PhysRevLett.83.2636}, and  Majorana chain  \cite{Kitaev_2001}.
The Uhlmann phase has been measured recently in an interacting system involving mixed states, which consists of a  topological insulator simulated by entangled superconducting qubits \cite{viyuelanpj18}.
Also, alternative formulations of a geometric phase for mixed states 
were introduced by Sj\"oqvist   \textit{et al.} \cite{PhysRevLett.85.2845} in the context of quantum interferometry.  
This interferometric phase has been already verified in experiments involving nuclear magnetic resonance techniques \cite{PhysRevLett.91.100403}.  
Although the Uhlmann and interferometric phases have been experimentally verified, the physical interpretation of these two different geometric phases for mixed states is still open to investigation \cite{ole2016}.

A system that has been of interest to explore geometric phases, is that of two entangled spin-$\frac 1 2$ fermions, where one of the spins is driven by a time-dependent magnetic field. 
Several features of the Berry phase have been investigated in these models by considering different degrees of spin couplings \cite{XING200654,SUN2007119,yiprl04}.
In particular, some aspects regarding the additivity of the Berry phases of the composite system, and their corresponding subsystems, have been explored \cite{yiprl04,Hanprl08,PhysRevLett.100.168902}.
In the latter model, important aspects of the physics of two-qubit systems are recovered, which makes them interesting candidates to study the Uhlmann phases in entangled systems and its corresponding subsystems.
Another aspect that has not been  explored in this type of systems, is the role of quantum correlations \cite{Adesso2016,audre07} in the  structure of geometric phases and their corresponding topological nature.

In this work we investigate the Uhlmann phase of two interacting fermions of spin-$\frac 1 2$, driven by an on-site varying magnetic field. The degree of mixing of the system is described by a depolarizing channel model  \cite{nielsen11}. The aim of this work is two-fold:   to find and characterize the topological structure of the model under the effects of a uniform depolarization, and also explore  the role of  quantum correlations in the structure of the Uhlmann phase associated to entangled mixed states. 

The paper is organized as follows: 
in Sec.~\ref{subsec:spin} we present the model and discuss the procedure to calculate the Uhlmann phase. 
In Sec.~\ref{sec:compositesystemAB} we introduce the depolarization model to study the main features of the Uhlmann phase in a composite system for entangled mixed states. 
In Sec.~\ref{sec:subsysAandB}  we derive an exact analytical expression for the Uhlmann phase in the corresponding subsystems. The topological structure is explored for different depolarization strengths, and its relation with the concurrence of the system is also investigated. 
Finally,  in Sec.~\ref{sec:conclusions} we present the conclusions.

\section{Model}\label{subsec:spin}

A system  of two interacting fermions of spin-$\frac{1}{2}$, driven by a time-dependent magnetic field  $\bm{B}(t)$,  can be described by a Hamiltonian \cite{yiprl04}
\begin{equation}
\hat{H}_0=\frac 1 2  \,\bm{\sigma}_1\cdot\bm{B}(t)+J\left(\sigma_1^+\sigma_2^+ + \sigma_2^-\sigma_1^-\right),
\label{hamyirotado}  
\end{equation}
%
%
where $\bm{\sigma}_j=(\sigma_j^x,\sigma_j^y,\sigma_j^z)$, and 
%
$\sigma^{\pm}_j=(\sigma_j^x \pm i\sigma_j^y)/2$.
The rotating magnetic field $\bm{B}(t)=B_0\, \hat{\bm{n}}(t)$ is acting on one of the spins along the direction $\hat{\bm{n}}(t)=(\sin\theta \cos\phi,\sin\theta\sin\phi,\cos\theta)$. The azimuthal direction is $\theta\in[0,\pi]$, and we assume that the time dependence is on the parameter $\phi=\phi(t)$, with $\phi\in[0,2\pi]$.
%
%
%
The eigenvalues, $E_j$, of the rescaled Hamiltonian $\hat{H}=\hat{H}_0/(B_0/2)$  in the composite spin basis $\{\ket{+-},\ket{++},\ket{--}, \ket{-+} \}$, are given by,
\begin{eqnarray}
E_1=-E_2=\sqrt{1+g^2/2+g\sqrt{g^2+4\sin^2\theta}/2}; \nonumber \\
E_3=-E_4=\sqrt{1+g^2/2-g\sqrt{g^2+4\sin^2\theta}/2},
\label{eigenvalues}    
\end{eqnarray}
where $g=2J/B_0$ measures the spin-spin coupling. 
The corresponding normalized eigenvectors are 
$\ket{u_j}={\cal N}_j^{-1/2}[u_j^{(1)}\,e^{-i \phi},u_j^{(2)},u_j^{(3)},u_j^{(4)} \,e^{i \phi}]^T$,
where
\begin{equation}
\begin{aligned}
u_j^{(1)}&= \sin\theta;
&\qquad
u_j^{(2)}&=g\,\left[\frac{\cos^2\theta-E_j^2}{1-E_j^2}\right]; 
\\
u_j^{(3)}&=E_j-\cos\theta; 
& 
u_j^{(4)}&=g\sin\theta\,\left[\frac{\cos\theta-E_j}{1-E_j^2}\right],  
\label{constants}
\end{aligned}
\end{equation}
with ${\cal N}_j=\sum_i\left[u_j^{(i)}\right]^2$.
The \textit{ground state} of the system is $\ket{u_2}$, with an energy $E_2$. 
The instantaneous eigenstates $\ket{u_j}$ 
of the composite spin-$\frac 1 2 $ system can be seen as vectors in an enlarged Hilbert space ${\cal H}_{AB}={\cal H}_A \otimes  {\cal H}_B$, 
where subsystem $A$ corresponds to the spin where the magnetic field is applied.
This allows to define a density matrix for each composite $j$-th state as $\rho=\ket{u_j}\bra{u_j}$.
This feature of the model has allowed to explore geometric phases, such as the Berry phase, associated to subsystem-subsystem coupling \cite{PhysRevLett.85.2845,yiprl04,SUN2007119}. 
An  approach to explore the geometric phases in composite systems is by using the Uhlmann phase \cite{uhlmannrmp86,uhlmannlmp91}.  %
The Uhlmann phase, $\Phi$, introduced by Viyuela \cite{viyuelaprl14,Viyuela_2015} for exploring thermal effects in one-dimensional fermion systems is given by
\begin{equation}
\Phi={\rm Arg}\left\{{\rm Tr}[\rho_{\lambda_0}\,V(\lambda,\lambda_0)]\right\},
\label{uhlmana}
\end{equation}
where $V(\lambda,\lambda_0)={\cal P} e^{\oint A(\lambda)}$ is a  $\lambda$  ordered integral, and $A(\lambda)$ is the Uhlmann connection. In general $A(\lambda)$ does not commute for all values of the parameter $\lambda$. 
An alternative procedure to evaluate $V(\lambda,\lambda_0)$ is by solving the differential equation for the evolution operator,
\begin{equation}
d V(\lambda,\lambda_0)/d\lambda=A(\lambda) \,V(\lambda,\lambda_0),
\label{difqeV}
\end{equation}
with the initial condition $V(\lambda_0,\lambda_0)=\mathbb{1}$. We assume that $\lambda_0=\lambda(t=0)$.
The Uhlmann connection $A(\lambda)$ is given by:
\begin{equation}
A(\lambda)=\sum_{i,j}\ket{\psi_i}\frac{\braket{\psi_i|\left[\partial_{\lambda}\sqrt{\rho}, \sqrt{\rho}\right]|\psi_j} }{p_j+p_i} \bra{\psi_j}\,d\lambda,
\label{uhlAU}
\end{equation}
which involves the matrix elements with respect to the eigenbasis $\{\ket{\psi_j}\}$ of the density matrix $\rho$, which we assume to be diagonalized, with eigenvalues $\{p_j\}$. In the spectral basis, $\rho=\sum_{j}p_j\ket{\psi_j}\bra{\psi_j}$.
By explicitly  computing the matrix elements of the commutator in Eq.~(\ref{uhlAU}) we write the Uhlmann connection as, 
\begin{equation}
A(\lambda)=\sum_{i\ne j}\frac{(\sqrt{p_j}-\sqrt{p_i})^2}{p_j+p_i}\braket{\psi_i|\partial_{\lambda}\psi_j} \ket{\psi_i}\bra{\psi_j}\,d\lambda.
\label{uhlAUb}
\end{equation}

In the next sections we investigate the general features of the Uhlmann phase in mixed entangled states for a composite system $AB$ (Sec.~\ref{sec:compositesystemAB}), and its corresponding subsystems $A$ and $B$ (Sec.~\ref{sec:subsysAandB}).
The degree of mixing of each state is provided by a \textit{depolarizing channel} model. 
The latter is used to explore the underlying mechanism that allows to control the Uhlmann phase and its topological transitions, by tuning the depolarization strength.
Also, we study the role of quantum correlations in the topological transitions of the Uhlmann phase in  our two-qubit system.

\section{Uhlmann phase for a composite system with entangled mixed states}\label{sec:compositesystemAB}

\begin{figure}[H] 
\begin{center}
\begin{subfigure}{4.7cm}
\includegraphics[width=4.5cm]{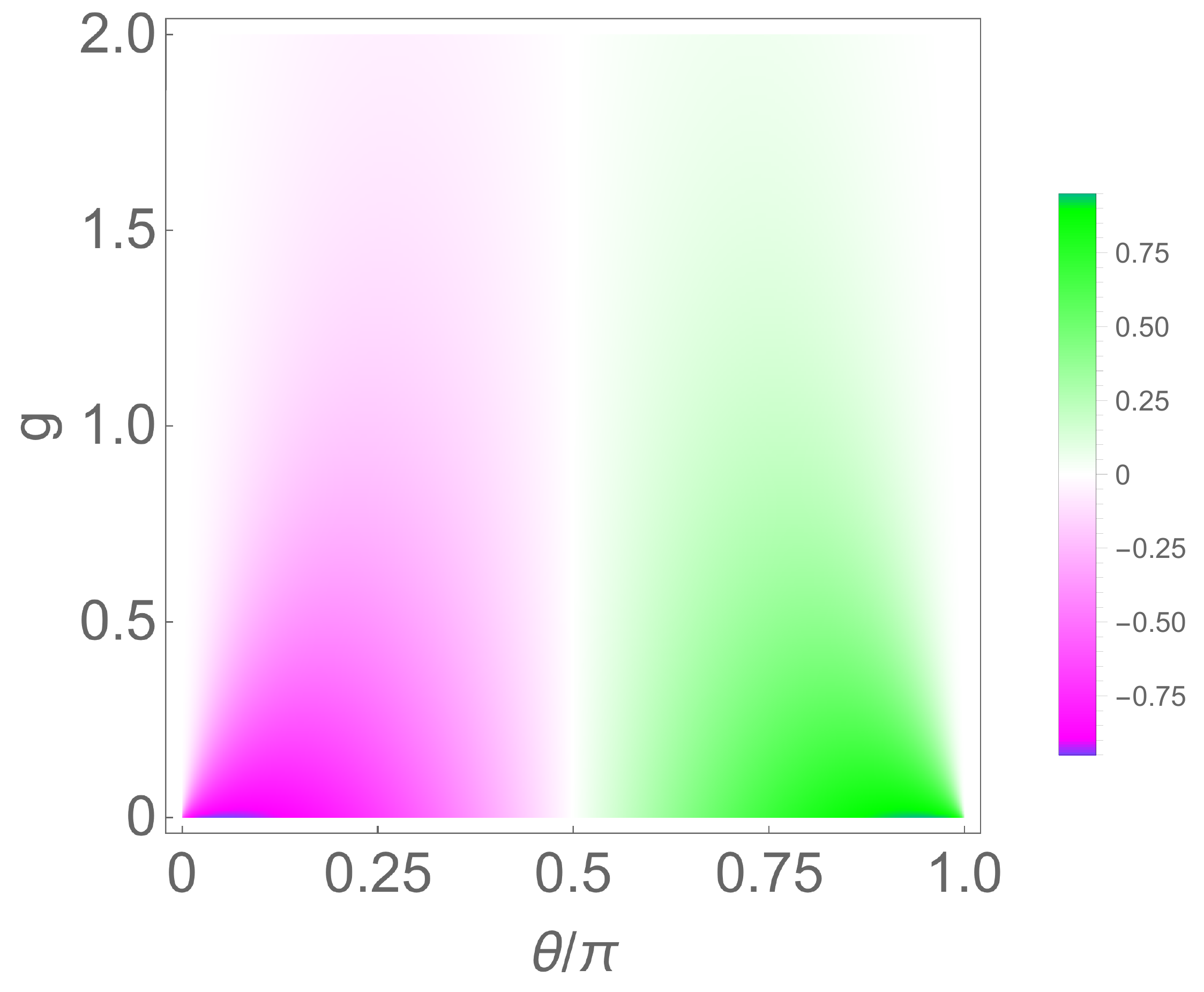}
\caption{Uhlmann phase $\Phi_{d}(q_1)$. }
\end{subfigure}%
\begin{subfigure}{4.7cm}
\includegraphics[width=4.5cm]{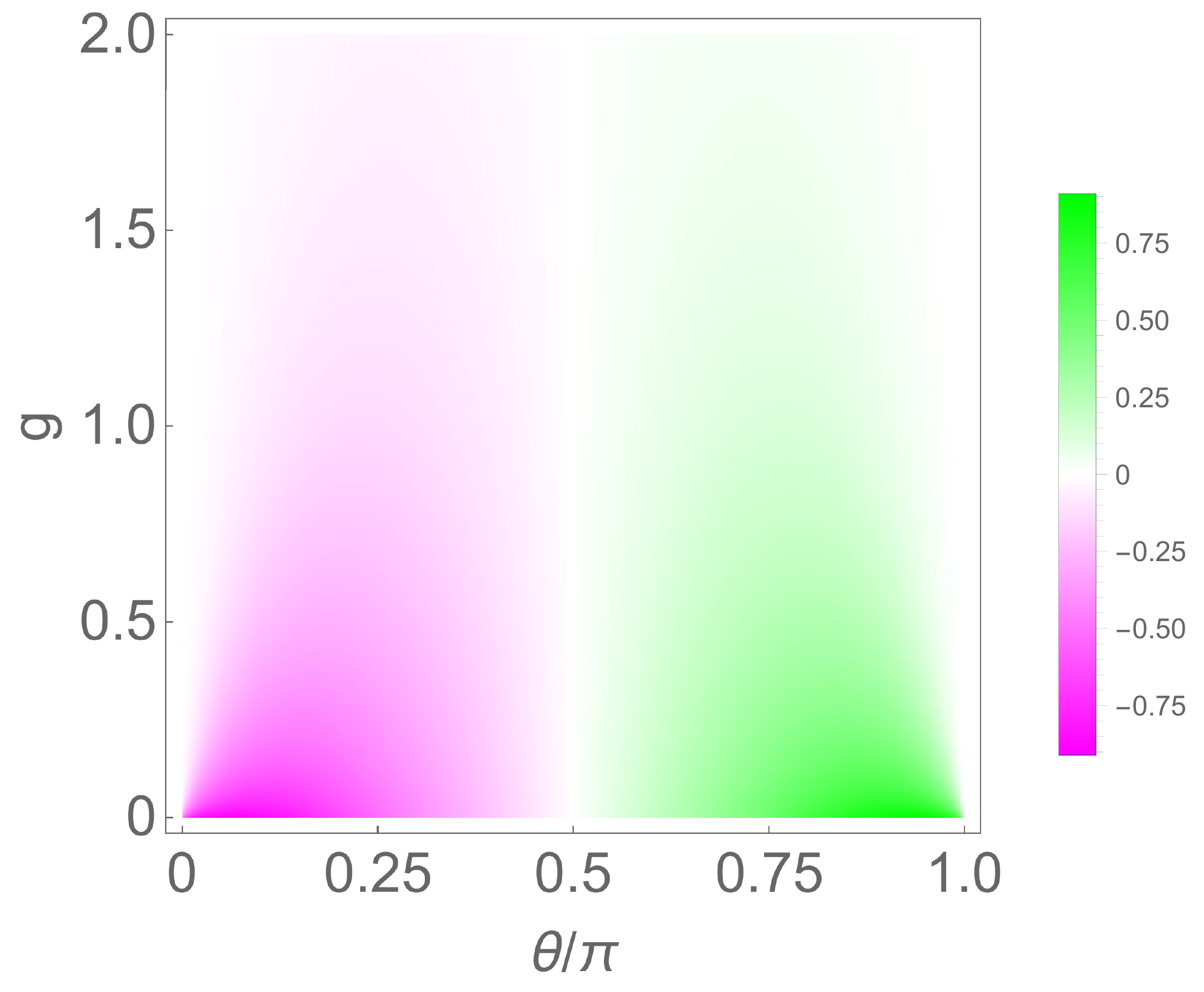}
\caption{Uhlmann phase $\Phi_{d}(q_2)$.  }
\end{subfigure}
\begin{subfigure}{4.7cm}
\includegraphics[width=4.5cm]{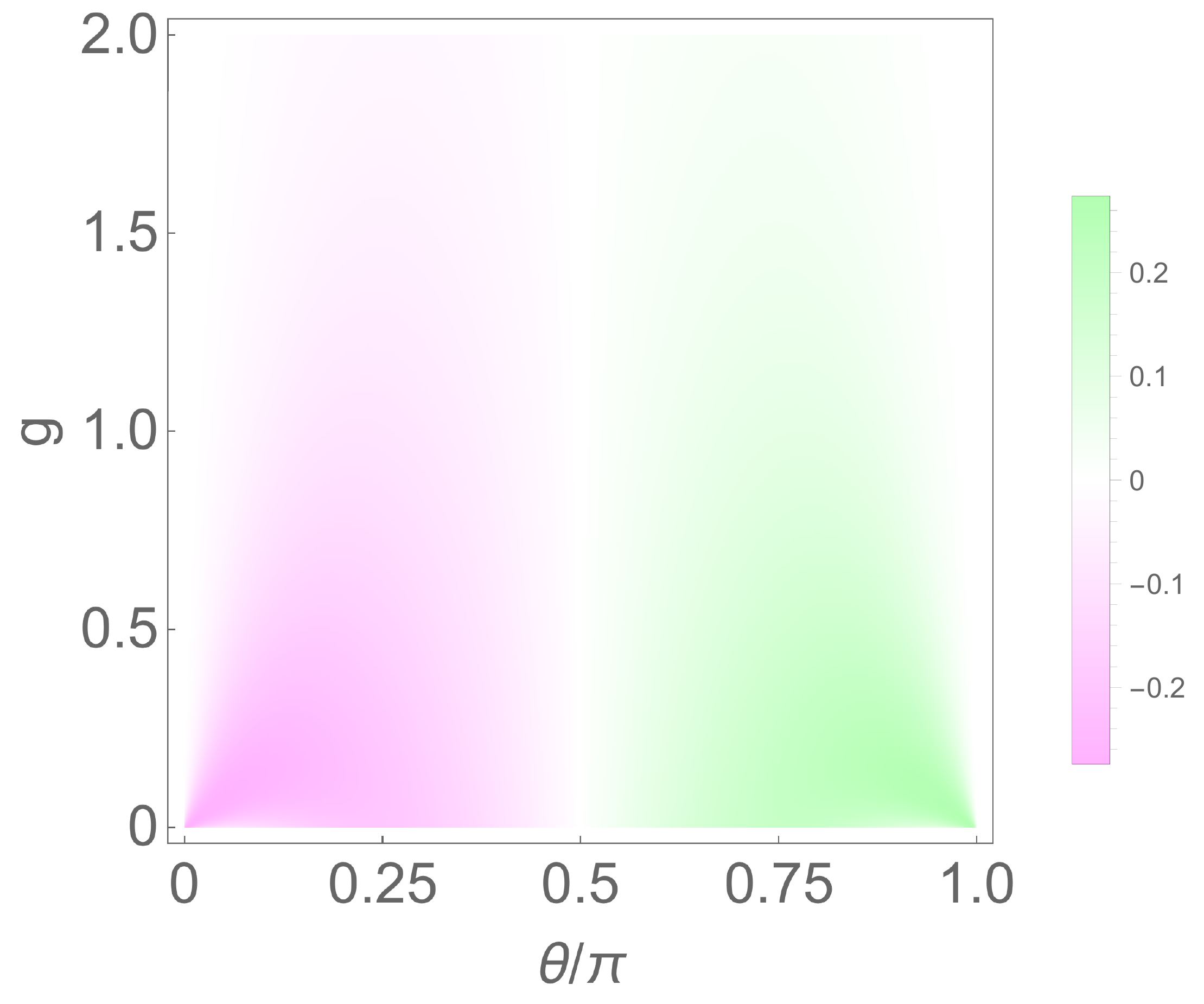}
\caption{Uhlmann phase $\Phi_{d}(q_3)$.}
\end{subfigure}%
\begin{subfigure}{4.7cm}
\includegraphics[width=4.5cm]{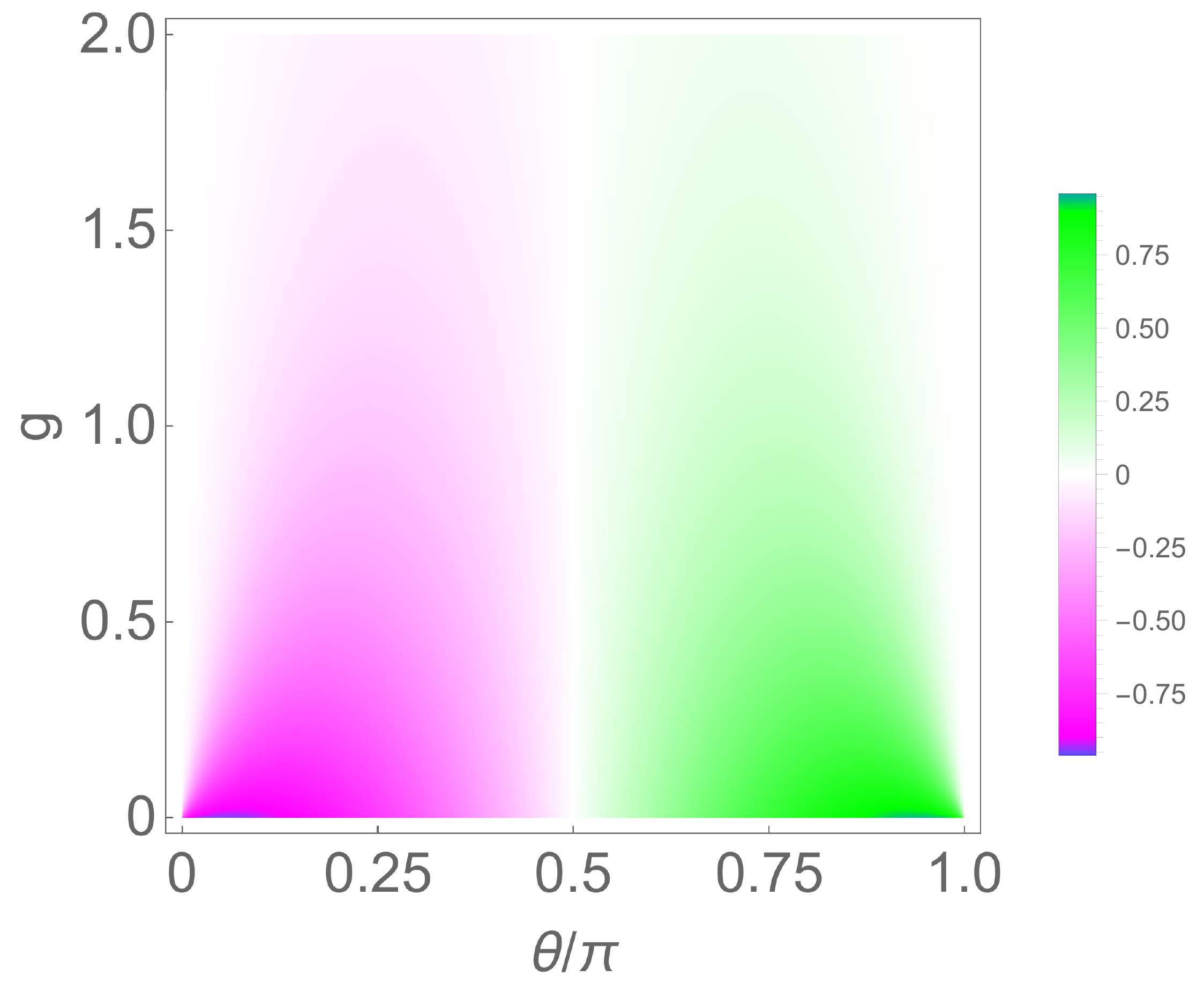}
\caption{Berry phase $\gamma$.}
\end{subfigure}%
\end{center}
\caption[]{(a)-(c) Color density maps of the Uhlmann phase of the composite system, $\Phi_{d}(q)$, for different values of the depolarization strength, $q$, as a function of the coupling parameter $g$, and $\theta$. We consider the following values of $q$: (a) $q_1=0.01$, (b) $q_2=0.1339$, and  $q_3=0.25$.  In (d) we include    the Berry phase of the ground state of the composite system $\gamma$ [Eq.~(\ref{Berry_uj})]. Observe that the latter coincides with the Uhlmann phase of case (a) for very low values of the polarization strength, as expected.  Here, and in all the paper we choose the \textit{ground state} of the system ($j=2$). Also, all the phases are given in units of $\pi$.}
\label{colormap1ABDepolarizing}
\end{figure}

An appropriate model to investigate the features of the Uhlmann phase for entangled mixed states is that of a \textit{depolarizing channel} \cite{nielsen11}, which allows to explore these entangled mixed states under uniform depolarization.
We consider a depolarizing channel  
described by the density matrix of the composite system, $AB$,
\begin{equation}
\rho_{d}=(q/4)\,\mathds{1}_4 +(1-q)\,\rho, \label{werenerdensity}
\end{equation}
where $\rho= \ket{u_j}\bra{u_j}$ is the density matrix of the $j$-th pure state,  an $q$ is the strength of the depolarization, with $q \in [0,1]$.
Since $\rho_{d}$ and $\rho$ share the same eigenbasis $\{\ket{u_j}\}$, we can use the completeness relation 
$\sum_{k} \ket{u_k}\bra{u_k}=\mathds{1}_4$,
to obtain the eigenvalues of $\rho_{d}$ involving the $j$-th state, given by 
$p_{d,k}=q/4+(1-q)\,\delta_{k,j}$.
To calculate the Uhlmann connection $A(\lambda)$ given by Eq.~(\ref{uhlAUb}), we fix the notation in the Uhlmann procedure  to suit our problem, by letting the eigenstates $\ket{\psi_j}\rightarrow \ket{u_j}$, and the parameters $\lambda\rightarrow\phi$, and $\lambda_0\rightarrow\phi_0$.
The  terms $\braket{u_i|\partial_{\phi}u_j}$ in (\ref{uhlAUb}) can be %
evaluated explicitly, and yield 
%
%
$\braket{u_i|\partial_{\phi}u_j}=i\,(u_i^{(4)}u_j^{(4)}-u_i^{(1)}u_j^{(1)})/\sqrt{{\cal N}_i\,{\cal N}_j}$.
Thus, the Uhlmann connection $A_{d}(\phi)$ for the depolarization channel case is, 
\begin{eqnarray}
A_{d}(\phi)&=&\sum_{i\ne j}\frac{i}{\sqrt{{\cal N}_i\,{\cal N}_j}}\frac{\left(\sqrt{p_{d,j}}-\sqrt{p_{d,i}}\right)^2}  {p_{d,j}+p_{d,i}} \times \nonumber  \\
&& \left(u_i^{(4)}u_j^{(4)}-u_i^{(1)}u_j^{(1)}\right) \ket{u_i}\bra{u_j}\,d\phi.
\label{uhlAUbbis2}
\end{eqnarray}
Using Eq.~(\ref{uhlmana}) we obtain the Uhlmann phase, $\Phi_{d}$, for the depolarizing channel case,  
by numerically computing  $V(\phi,\phi_0)$ from (\ref{difqeV}), which satisfies $dV(\phi,\phi_0)/d\phi=A_{d}(\phi)V(\phi,\phi_0)$
with the initial condition $V(\phi_0,\phi_0)=\mathds{1}_4$. 
For comparison,  we also consider  
the  Berry phase, $\gamma$, corresponding to each of the eigenstates, $\ket{u_j}$, 
\begin{equation}
\gamma=\int_0^{2 \pi}d\phi\,\braket{u_j|i\partial_{\phi}u_j}=\frac{2\pi}{{\cal N}_j}\left\{\left[u_j^{(1)}\right]^2-\left[u_j^{(4)}\right]^2\right\}.
\label{Berry_uj}    
\end{equation}
In Fig.~\ref{colormap1ABDepolarizing} we show the behavior of the geometric phases for the composite system $AB$ as a function of the coupling parameter $g$ in all directions of the field, for different values of the depolarization strength, $q$.

\begin{figure}[H]
 \begin{center}
   {\includegraphics[angle=0,width=3.5in]{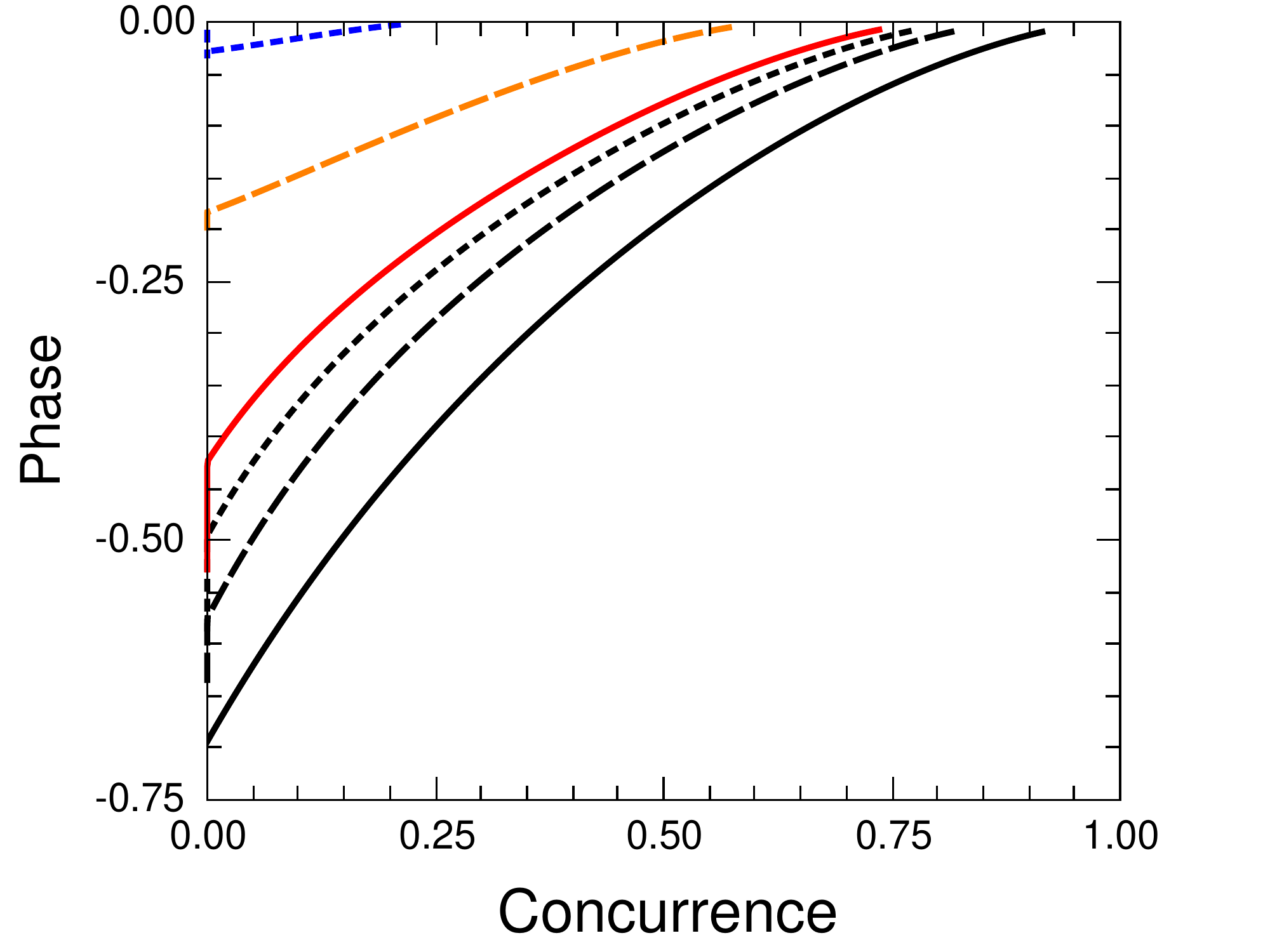}}
    \end{center}
      \caption {Parametric plot of the Uhlmann phase, $\Phi_{d}(g)$, and the concurrence, ${\cal C}(\rho_d)(g)$, as a function of the coupling parameter $g\in[0.01,4]$, for a fixed direction $\theta=\pi/4$. We consider different values of the depolarization strength, $q$: 0.01 (black solid line), 0.07508 (black dashed line), 0.107321 (black dotted line), 0.1339 (red solid line), 0.25 (orange dashed line), and 0.5 (blue dotted line).}
            \label{parametric1}
             \end{figure}

In Figs.~\ref{colormap1ABDepolarizing}(a)-(c) we show that the Uhlmann phase $\Phi_d$ is 
non trivial, with an evanescent magnitude as the polarization strength $q$ is increased. In Fig.~\ref{colormap1ABDepolarizing}(d) we include the Berry phase $\gamma$  [Eq.~(\ref{Berry_uj})] for the \textit{ground state}, and show that the latter agrees with the Uhlmann phase for very small values of $q$ case, as depicted in Fig.~\ref{colormap1ABDepolarizing}(a).
This is an expected result because in the limit $q\rightarrow 0$, the density matrix of the composite system tends to the density matrix of a pure state \textit{i.e.} $\rho_{d}\rightarrow \rho$. 
Recently, the degree of mixing in the subsystems associated to a pure state has become relevant in the Uhlmann procedure for a two--band model of topological systems simulated by a qubit \cite{viyuelanpj18}.  
Although the mixing is important to describe geometric phases, we wonder about the role of quantum correlations in models with interacting fermions with spin.  
In our particular model, we are dealing with a geometric phase that has information about the inherent non-classical correlations induced by the spin-spin coupling.
A relevant quantity that measures the degree of \textit{entanglement} in  quantum systems is the \textit{concurrence} \cite{wooprl98,wooqic01,audre07},  ${\cal C}(\rho)$, defined as,
\begin{equation}
{\cal C}(\rho)={\rm max}\{0,\lambda_1-\lambda_2-\lambda_3-\lambda_4 \},
\label{concurrence}
\end{equation}
where $\lambda_i$ ($i=1,2,3,4$) are the square roots of the eigenvalues of $\rho\, \bar{\rho}$,  in descending order.
Here, $\bar{\rho}$ stands for the result of applying a spin-flip operation to $\rho$ \textit{i.e.} $\bar{\rho}=(\sigma_y \otimes\sigma_y)\,\rho^* \,(\sigma_y \otimes\sigma_y)$, using the Pauli matrix $\sigma_y=\big(\begin{smallmatrix}
0  & - i  \\
i  &  0
\end{smallmatrix}\big)$.
In Fig.~\ref{parametric1} we show a parametric plot of the Uhlmann phase versus the concurrence of the composite system as a function of the coupling parameter $g$, for a fixed direction $\theta$.

We show that in general, the geometric phase, $\Phi_{d}(g)$, and the concurrence  ${\cal C}(\rho_d)(g)$, exhibit a monotonically decreasing behavior under a continuous variation of the parameter $g$, for the  different values of the depolarization strength.  
In view of the above results, we argue that there must be a deeper connection between the geometric phase and the concurrence of the system.
We shall explore this issue in more detail in the next section, where we address the problem of geometric phases for subsystems $A$ and $B$.

\section{Uhlmann phase for the subsystems.}\label{sec:subsysAandB}

Our aim in studying the Uhlmann phase for subsystems $A$ and $B$ is threefold:  to understand how this geometric phase changes under uniform depolarization; 
to determine the role played by the quantum correlations of the two-qubit entangled system in the formation process of these phases, 
and to explore to what extent these measurements of entanglement allow to characterize their  topological phase transitions.

\subsection{ Depolarizing channel with $q=0$ (pure state) }\label{subsec:depq0}

We study the geometric phase of subsystems $A$ and $B$ 
for the case $q=0$ associated to a \textit{pure state}, that is, $\rho_{d}=\rho$,  and  investigate the main features of  the Uhlmann and Berry phases.
The density matrices for the subsystems $A$ and $B$ corresponding to the $j$-th eigenstate, are obtained by computing the trace of $\rho$ over $A$ (or $B$) \textit{i.e.}  $\rho^A={\rm Tr}_B[\rho]$, and  $\rho^B={\rm Tr}_A[\rho]$, respectively.
The density matrices for the subsystems are represented by the general $2 \times 2$ matrices,
\begin{equation}
\rho^s=
\begin{pmatrix}
a_{s} & c_{s} \,e^{-i \phi}\\ 
c_{s}  \,e^{+i \phi}  & 1-a_{s}
\end{pmatrix}, 	
\label{bastardnotation}
\end{equation}
where the real coefficients $a_{s}$, and $c_{s}$ ($s=A,B$) for each eigenstate,  depend both on the direction $\theta$, and the coupling parameter $g$, but are independent on the parameter $\phi$:
\begin{eqnarray}
a_{A}(\theta,g)&=&{\cal N}_j^{-1}\left[ \left(u^{(1)}_j\right)^2+\left(u^{(2)}_j\right)^2\right]; \nonumber \\ 
c_{A}(\theta,g)&=&{\cal N}_j^{-1}\left[ u^{(1)}_j\,u^{(3)}_j+ u^{(2)}_j\,u^{(4)}_j\right],
\label{coefA}
\end{eqnarray}
and
\begin{eqnarray}
a_{B}(\theta,g)&=&{\cal N}_j^{-1}\left[ \left(u^{(1)}_j\right)^2+\left(u^{(3)}_j\right)^2\right]; \nonumber \\ 
c_{B}(\theta,g)&=&{\cal N}_j^{-1}\left[ u^{(1)}_j\,u^{(2)}_j+ u^{(3)}_j\,u^{(4)}_j\right].
\label{coefB}
\end{eqnarray}
The eigenvalues of $\rho^s$  are
\begin{eqnarray}
p_{s,1}&=&\left[1-\sqrt{(1-2a_{s})^2+4c_{s}^2}\, \right]/2; \\
p_{s,2}&=& \left[1+\sqrt{(1-2a_{s})^2+4c_{s}^2}\, \right]/2,
\label{eigenvalspin}    
\end{eqnarray}
which satisfy the conditions  $p_{s,1}+p_{s,2}=1$, and $p_{s,1}\, p_{s,2}={\rm det}[\rho^s]=a_{s}(1-a_{s})-c_{s}^2$. The corresponding eigenvectors are,
\begin{equation}
\ket{v_{s,l}}=\frac{1}{\sqrt{N_{s,l}}}
\begin{bmatrix}
\beta_{s,l}  \,e^{-i \phi}\\ 
1 
\end{bmatrix}, \\
\label{eigenvecspin}    
\end{equation}
where $l=1,2$,  $N_{s,l}=\beta_{s,l}^2+1$,
with $\beta_{s,l}=c_{s}/(p_{s,l}-a_{s})$.
The Uhlmann connection  can be computed from Eq.~(\ref{uhlAUb}), by considering the variation of the parameter $\phi$,
which leads us to $A^s(\phi)=-2i\Delta p_{s}\,(\bm{n}_{\delta_s}\cdot \bm{\sigma})\, d\phi$,  with $\bm{n}_{\delta_s}=(-\delta_s\cos\phi,-\delta_s\sin\phi,1)$, where
$\Delta p_{s}=(\sqrt{p_{s,2}}-\sqrt{p_{s,1}})^2/N_{s,1}N_{s,2}$, 
and the parameter $\delta_s=(2a_{s}-1)/2c_{s}$.
The Uhlmann connection of the subsystems can be written in terms of the \textit{concurrence} of the composite system, ${\cal C}(\rho$), which 
for the particular case of a \textit{pure state},
can be represented in term of the density matrix $\rho^s$ of the subsystems  as follows \cite{PhysRevA.91.032327}   
\begin{equation}
{\cal C}(\rho)=\sqrt{2\,(1-{\rm Tr}[(\rho^s)^2])},
\label{concurrencealternabis}
\end{equation}
which also can be expressed as ${\cal C}(\rho)=2\sqrt{{\rm det}[\rho^s]}=2\sqrt{p_{s,1}p_{s,2}}$.
Therefore, the factor $\Delta p_{s}=[1-{\cal C}(\rho)]/N_{s,1}N_{s,2}$,
appearing in the connection  $A^s(\phi)$, shows an explicit connection to the concurrence of the composite system.

We derive an exact analytical solution for the \textit{Uhlmann phase}, $\Phi^s$, of  subsystem $s$, associated to the $j$-th eigenstate by 
following a procedure that involves the explicit calculation of the evolution operator in a rotating frame \cite{Bohm}. The procedure yields the following Uhlmann phase of the subsystems $A$ and $B$,
\begin{equation}
\Phi^s(\theta,g)={\rm Arg}\left\{-\cos(\pi r_s)-i\, \left[\bar{\gamma}^s-\pi\right]\, \frac{\sin(\pi r_s)}{\pi\,r_s} \right\},
\label{analiticalUhlmann}
\end{equation}
where $r_s$ is a quantity which can be written in terms of the Berry phases $\gamma^{s,l}$ of the eigenstates of the subsystem $s$ 
\begin{equation}
\gamma^{s,l}(\theta,g)=\int_0^{2\pi}\, d\phi \braket{v_{s,l}|i\partial_{\phi}v_{s,l}}=2\pi\, \left( \beta^2_{s,l}/N_{s,l}\right),
\label{BerrySubsystems2}    
\end{equation}
and of the concurrence ${\cal C}(\rho)$, in the form
\begin{equation}
r_s(\theta,g)= \left(1-\gamma^{s,1}\, \gamma^{s,2}\,\left[{1-\cal C}^2(\rho)\right]\,/\pi^2 \right)^{1/2}.
\label{lars}
\end{equation}
The result (\ref{analiticalUhlmann}) involves also the composed phase $\bar{\gamma}^s=\sum_{l=1}^{2} p_{s,l}\,\gamma^{s,l}$, for which 
it is verified that $\bar{\gamma}^A+\bar{\gamma}^B-2\pi= \gamma$.
Note also that the factor in square brackets in Eq.~(\ref{lars}) is 
the radius $R$ of the Bloch sphere associated to the density matrix $\rho^s=(\mathbb{1}_2+R\,\hat{\bm{n}}_s\cdot\bm{\sigma})/2$, 
with a polarization vector $\bm{n}_s=2c_{s}(\cos\phi,-\sin\phi, \delta_s)$, and $\hat{\bm{n}}_s=\bm{n}_s/\lVert\bm{n}_s \rVert$, that is,  $R\equiv\lVert\bm{n}_s \rVert=1-{\cal C}^2(\rho)$.
Although the weighted sum $\bar{\gamma}^s$ occurs naturally in the Uhlmann phase (\ref{analiticalUhlmann}), we do not argue that 
it is the appropriate phase of a mixture. Indeed, there exists some controversy about the use of $\bar{\gamma}^s$ to describe mixtures
 \cite{yiprl04,Hanprl08,PhysRevLett.100.168902}.
An alternative procedure to measure the Berry phase for entangled subsystems have been proposed \cite{Hanprl08}, which involves 
individual phase factors instead of phase angles only,
\begin{equation}
\gamma_I^s(\theta,g)={\rm Arg}\left\{\sum_{l=1}^{2} p_{s,l}\,e^{i\,\gamma^{s,l}}\right\}.
\label{BerrySubsystems}    
\end{equation}
This definition is consistent with the interferometric ($I$) phase defined by Sj\"oqvist \textit{et al.} in  Ref.~\onlinecite{PhysRevLett.85.2845}.

\begin{figure}[H] 
\begin{subfigure}{4.7cm}
\includegraphics[width=4.5cm]{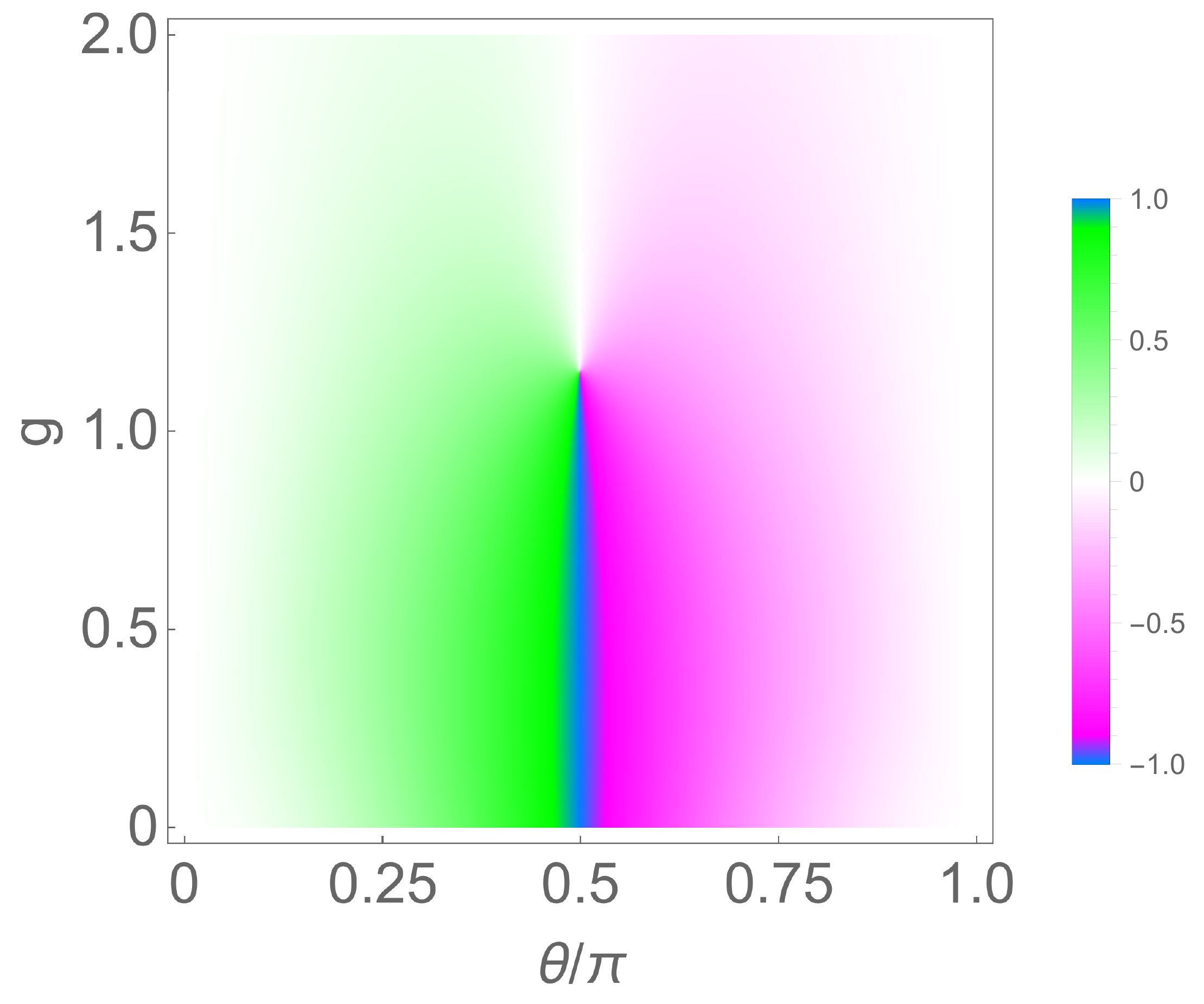}
\caption{Uhlmann phase $\Phi^A$. }
\end{subfigure}%
\begin{subfigure}{4.7cm}
\includegraphics[width=4.5cm]{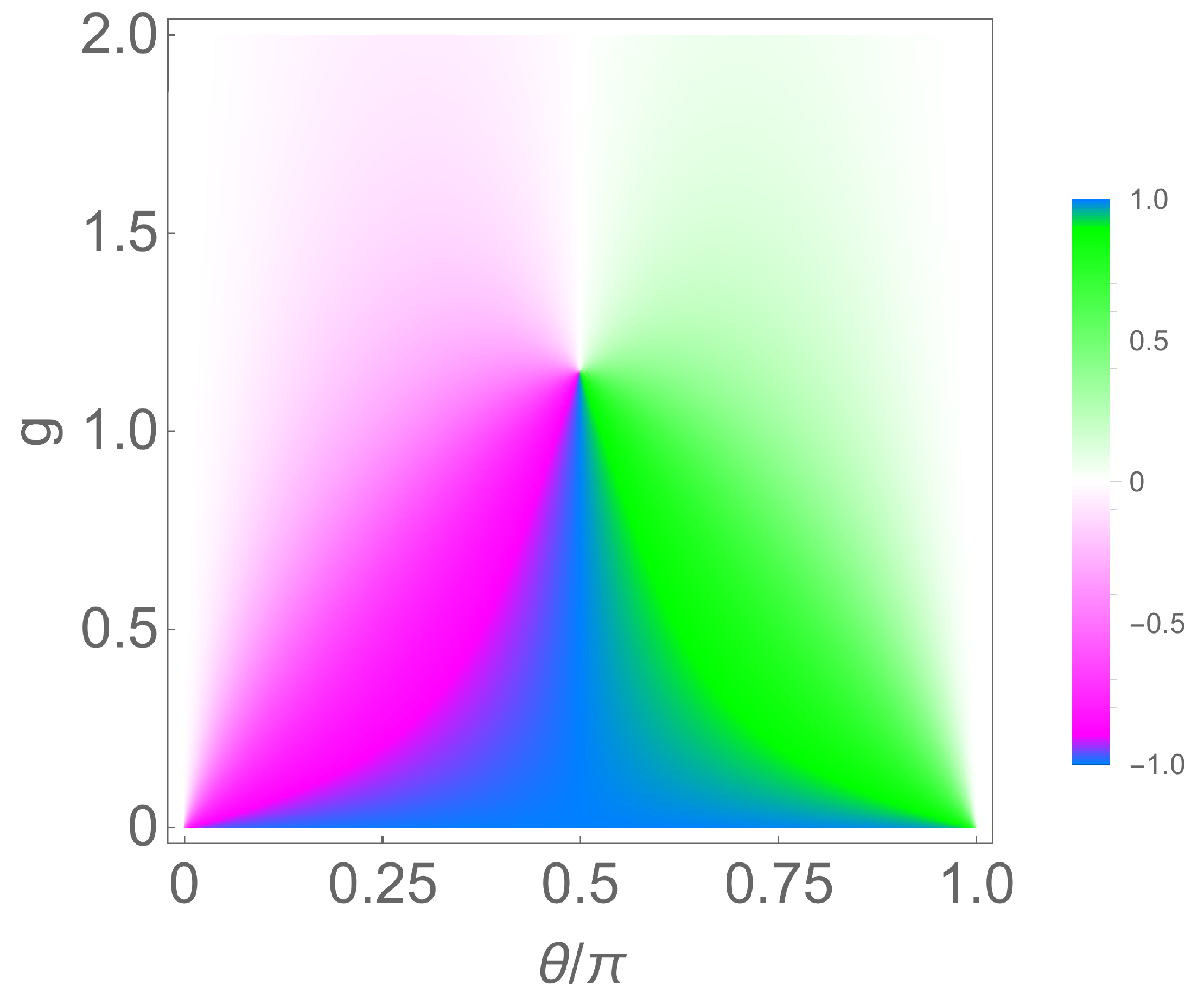}
\caption{Uhlmann phase $\Phi^B$.}
\end{subfigure}\vspace{10pt}
\begin{subfigure}{4.7cm}
\includegraphics[width=4.5cm]{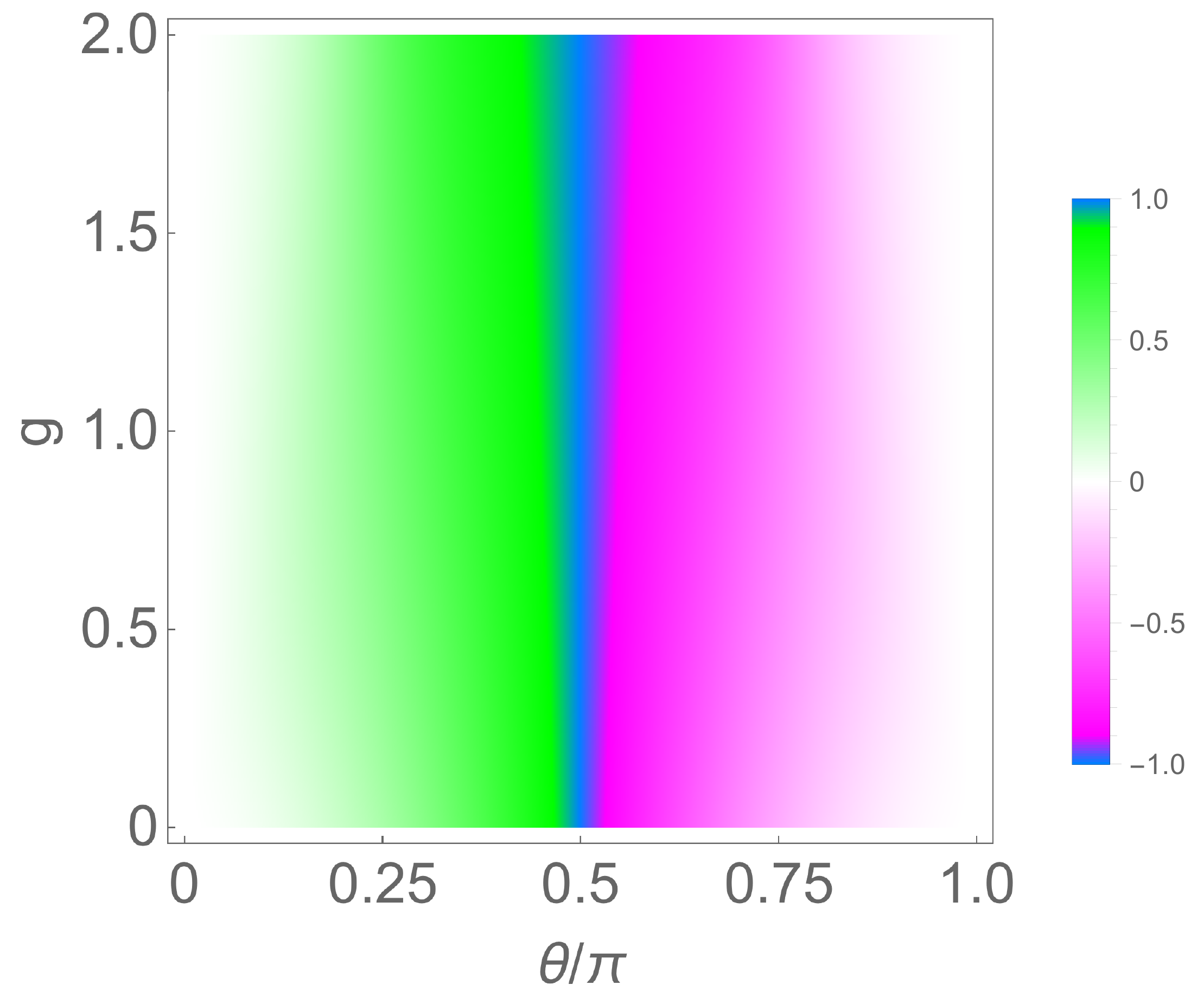}
\caption{Interferometric phase $\gamma_I^A$. }
\end{subfigure}%
\begin{subfigure}{4.7cm}
\includegraphics[width=4.5cm]{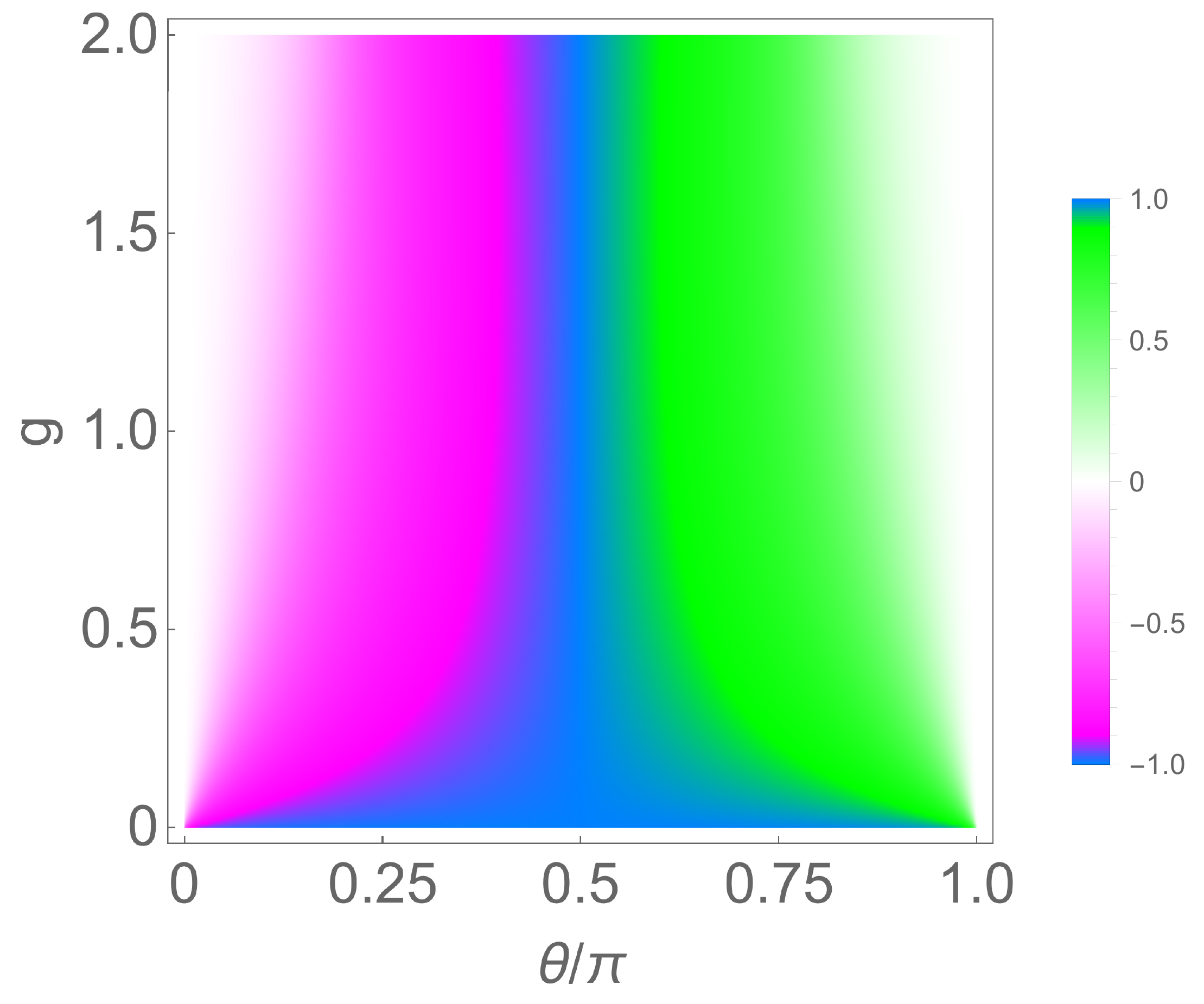}
\caption{Interferometric phase $\gamma_I^B$.}
\end{subfigure}\vspace{10pt}
\begin{subfigure}{4.7cm}
\includegraphics[width=4.5cm]{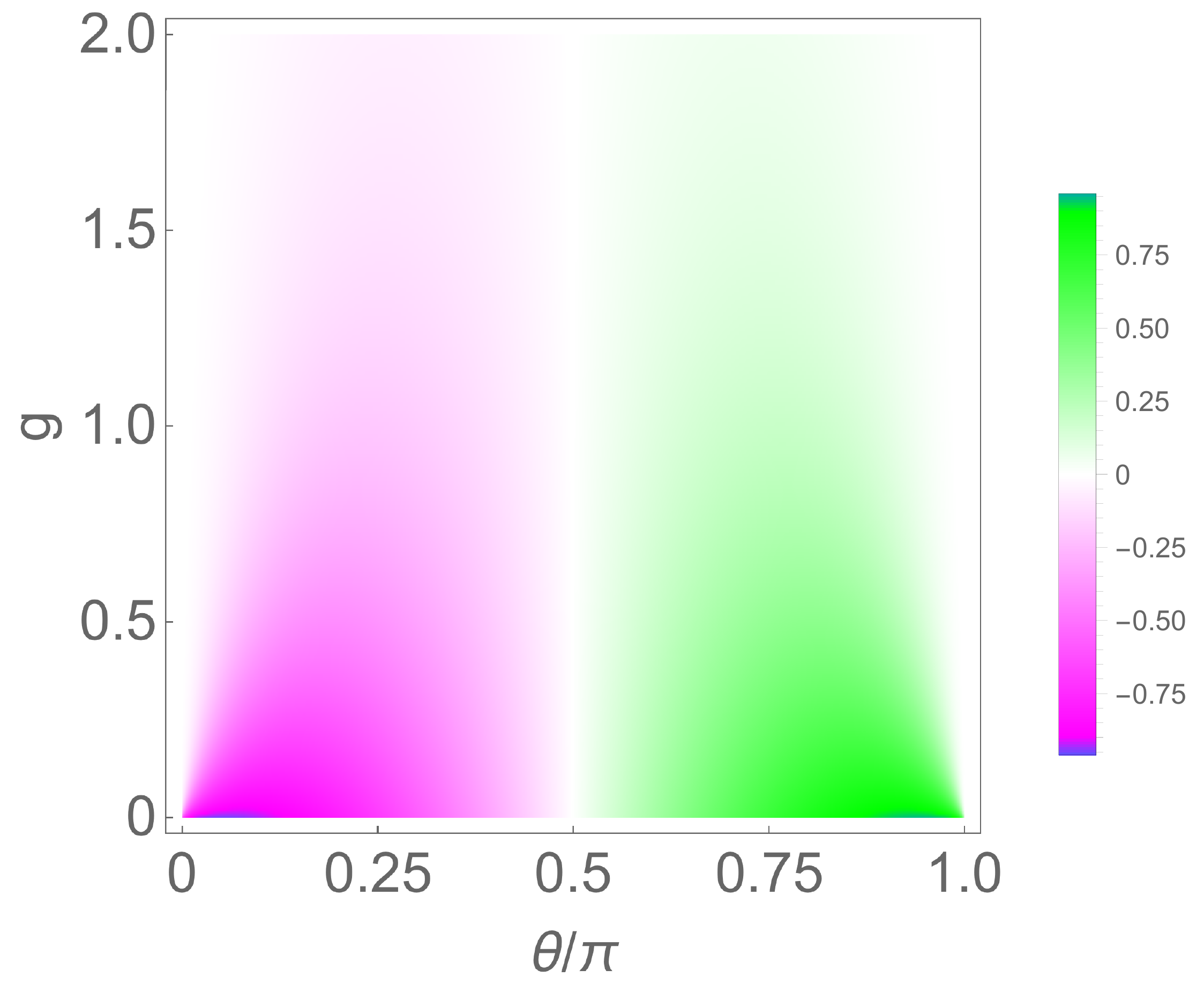}
\caption{Berry phase $\gamma$.}
\end{subfigure}%
\begin{subfigure}{4.7cm}
\includegraphics[width=4.5cm]{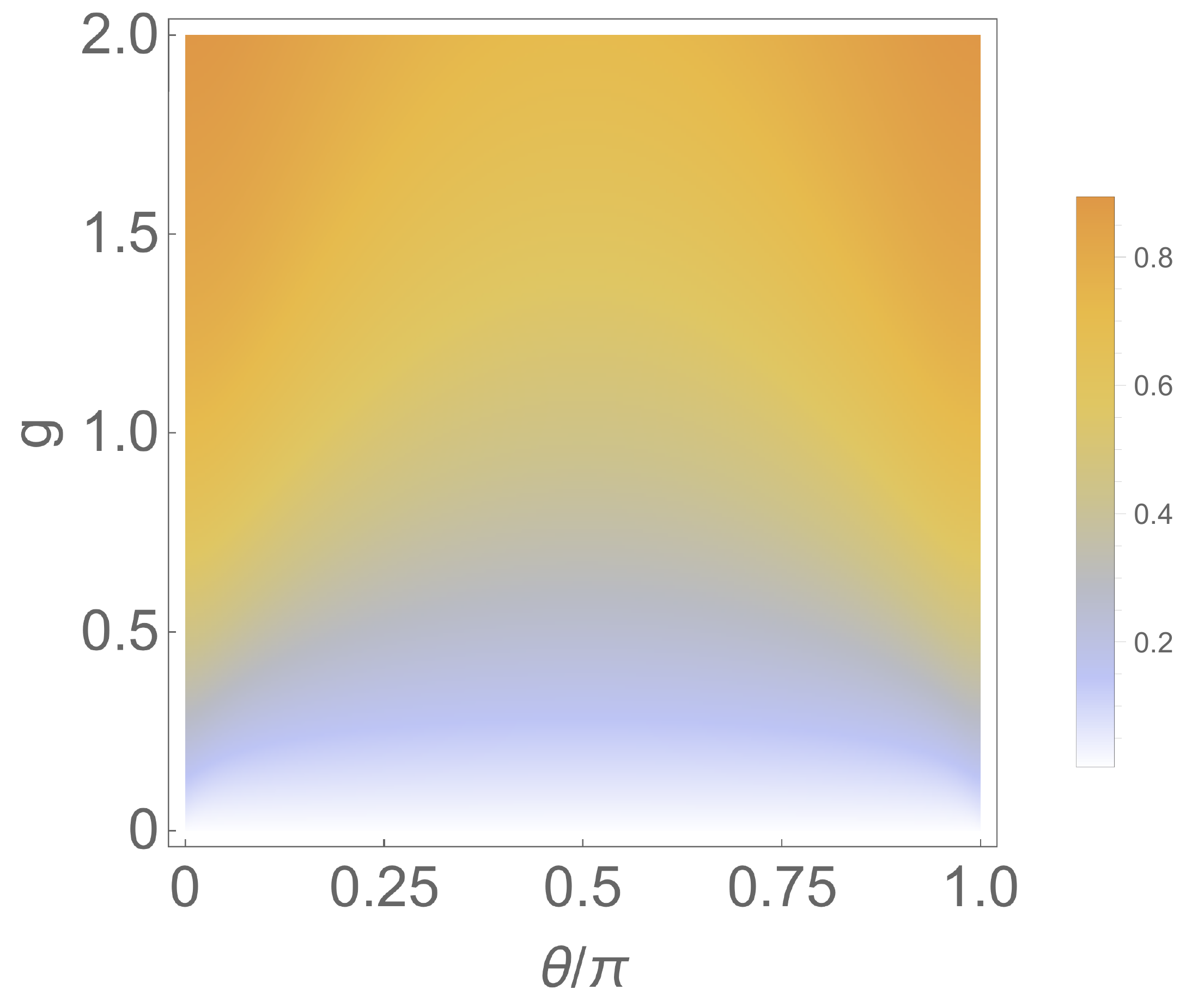}
\caption{Concurrence ${\cal C}(\rho)$.}
\end{subfigure}
\caption[]{Color density maps of (a)-(b) Uhlmann phase, $\Phi^s$ [Eq.~(\ref{analiticalUhlmann})] and (c)-(d)  interferometric phase, $\gamma_I^s$ [Eq.~(\ref{BerrySubsystems})] for the subsystems, as a function of the coupling parameter $g$, and $\theta$.
In cases (a)-(b) we emphasize on the presence of a vortex profile along $\theta=\pi/2$, occurring at a critical value of $g \simeq 1.15$. 
In (e) we include the Berry phase of the composite system,  $\gamma$ [Eq.~(\ref{Berry_uj})]. In (f) we present the concurrence of the composite system, ${\cal C}(\rho)$.}
\label{colormapjeq2}
\end{figure}

In what follows, we explore the features of the Uhlmann phase  $\Phi^s$ [Eq.~(\ref{analiticalUhlmann})], the interferometric phase [Eq.~(\ref{BerrySubsystems})], and their relationship with the Berry phase of the composite system, $\gamma$ [Eq.~(\ref{Berry_uj})].  
We  also investigate the effect of  quantum correlations on the behavior of these geometric phases.
In Figs.~\ref{colormapjeq2}(a) and (b) we present color density maps of the Uhlmann phase [Eq.~(\ref{analiticalUhlmann})] to show its dependence on the coupling, $g$, in all directions of the field. 
We show that $\Phi^s$ of the subsystems are in general different from each other. Nevertheless, they exhibit a distinctive feature along the direction $\theta=\pi/2$, which is a phase singularity (a ``phase vortex'') occurring at some critical value of the coupling, $g$, that characterizes a phase transition in the system. 
The phase is not defined at this point, and all colors meet there. Near such a point the whole range of values of the phase occurs.
We can appreciate in Figs.~\ref{colormapjeq2}(a) and (b) that for large values of the coupling, the Uhlmann phase $\Phi^s\rightarrow 0$.
In Figs.~\ref{colormapjeq2}(c) and (d) we have included the interferometric phase, $\gamma_I^s$  [Eq.~(\ref{BerrySubsystems})]. 
Note that the interferometric phase in Figs.~\ref{colormapjeq2}(c)-(d) does not exhibit a vortex structure, as in the case of the geometric phase, $\Phi^s$.
For small coupling $g$, we can appreciate that the Uhlmann phases shown in  Figs.~\ref{colormapjeq2}(a) and (b) converge, respectively, to the interferometric phases of Figs.~\ref{colormapjeq2}(c) and (d) that is,  $\Phi^s\rightarrow \gamma_I^s$.
In fact, by taking the limit  $g\rightarrow0$, one can show that for the \textit{ ground state} ($j=2$), $\Phi^A\rightarrow \gamma_I^A=\pi(1-\cos\theta)$, and $\Phi^B\rightarrow \gamma_I^B=\pi$.
We also observe in the low coupling $g$ regime that the behavior of $\Phi^s$ is very different in both subsystems. On the one hand, $\Phi^A$ increases from $0$ to $\pi$ as $\theta$ goes from $0$ to $\pi/2$, and from $-\pi$ to $0$ as $\theta$ goes from $\pi/2$ to $0$, with a clear phase transition occurring at $\theta=\pi/2$. While on the other hand, $\Phi^B$  also exhibits a phase transition at $\theta=\pi/2$, with a constant phase.
%
We have also included in Fig.~\ref{colormapjeq2}(e) the Berry phase $\gamma$ [Eq.~(\ref{Berry_uj})] of the composite system.
We show that in the limit of large values of the coupling, the Uhlmann phase of Fig.~\ref{colormapjeq2}(b) tends to  $\gamma$ of Fig.~\ref{colormapjeq2}(e), for all values of the direction $\theta$. Although it is difficult to compare with the color scale, we will address this point later by choosing different directions of the field. 
In Fig.~\ref{colormapjeq2}(f) we  also study the degree of entanglement in our \textit{two-qubit} coupled system, by exploring the behavior of  ${\cal C}(\rho)$.  
We can see that the small coupling regime, where the Berry and Uhlmann phases of the respective subsystems coincide, is characterized by a small concurrence (weak entanglement) of the ground state. 

Also from Fig.~\ref{colormapjeq2}(f) we may argue that in the large coupling regime, where the concurrence is high (strong entanglement), that $\Phi^B$ tends to the Berry phase of the composite system, $\gamma$, for all values of the direction $\theta$. 
Also, we can clearly appreciate that in the vicinity of the singularity at $\theta=\pi/2$, that ${\cal C}(\rho)\simeq1/2$.   
To gain more insight on the phase vortex, and  obtain the value of the critical coupling, let us consider the particular case of the direction $\theta=\pi/2$. 
In this case, the Uhlmann phases in  Eq.~(\ref{analiticalUhlmann}) can be simplified  because  $a_{s}=1/2$, $\delta_s=0$, and $\bar{\gamma}^s=\gamma^{s,l}=\pi$, which leads to $r_s={\cal C}(\rho)$.
Therefore, 
\begin{equation}
\Phi^{s}={\rm Arg}\left\{-\cos[\pi \,{\cal C}(\rho)]\right\},    
\label{UhlmansAyBpientre2}    
\end{equation}
which holds for all the eigenstates of the composite system.
Since ${\cal C}(\rho)\in[0,1]$, it follows from Eq.~(\ref{UhlmansAyBpientre2}) that the cosine function has only one node at  ${\cal C}(\rho)=1/2$ (or $R=3/4$), where the phase singularity occurs.
Also, from (\ref{concurrencealternabis}), we have found the following expression for the \textit{concurrence}  
\begin{equation}
{\cal C}(\rho)=g/\sqrt{g^2+4}.
\label{Cderho}    
\end{equation}
The result (\ref{Cderho}) allows to directly calculate the critical coupling, given by $g\equiv g_c=2/\sqrt{3}$, which corresponds to the 
position of the vortex of the Uhlmann phase observed in Figs.~\ref{colormapjeq2}(a)-(b).
\begin{figure}[H]
 \begin{center}
  {\includegraphics[angle=0,width=3.5in]{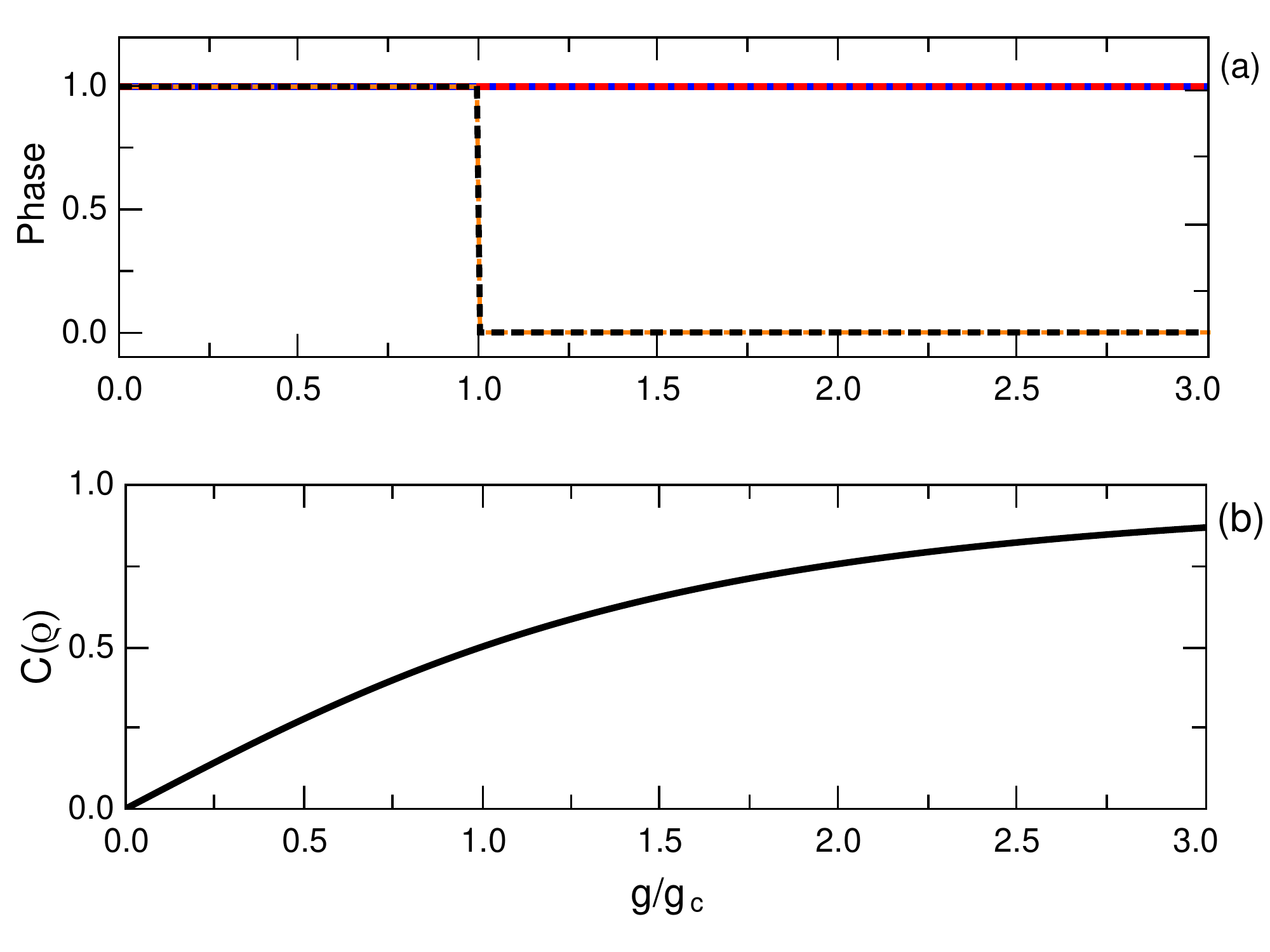}}
   \end{center}
     \caption {(a) Comparison of the geometric phases of the subsystems:  Uhlmann phase $\Phi^A$ (orange dashed line), and  $\Phi^B$ (black dotted line) [Eq.~(\ref{UhlmansAyBpientre2})], and interferometric phases [Eq.~(\ref{BerrySubsystems})], $\gamma_I^A$  (blue solid line), and $\gamma_I^B$ (red dotted line) as a function of the coupling parameter $g$.  (b) Concurrence, ${\cal C}(\rho)$, for the $j$-th eigenstate of the composite system (black solid line). The four eigenstates share the same concurrence. The direction $\theta=\pi/2$ was chosen for this calculation. }
          \label{subsyAyBpientre2_allj}
           \end{figure}
In Fig.~\ref{subsyAyBpientre2_allj} we show  $\Phi^{s}$ [Eq.~(\ref{UhlmansAyBpientre2})] as a function of the coupling parameter. We show that for small values of $g$, the Uhlmann phase equals to $+\pi$, until a node is reached where it jumps to $0$. This topological transition from a non-trivial to a trivial phase occurs at values of $g/g_c=1$.
The interferometric phases of the subsystems [Eq.~(\ref{BerrySubsystems})] yield $\gamma_I^s={\rm Arg}[-1]=+\pi$, and are also included in Fig.~\ref{subsyAyBpientre2_allj}. 
The Berry phase of the composite system [Eq.~(\ref{Berry_uj})] for each eigenstate $\ket{u_j}$ yields $\gamma=0$.  
From the results observed in Figs.~\ref{colormapjeq2} and \ref{subsyAyBpientre2_allj}, we emphasize on the relevance of the concurrence of the composite system, ${\cal C}(\rho)$, in the global behavior of  the Uhlmann phase, $\Phi^s$. 

\begin{figure}[H]
 \begin{center}
  {\includegraphics[angle=0,width=3.5in]{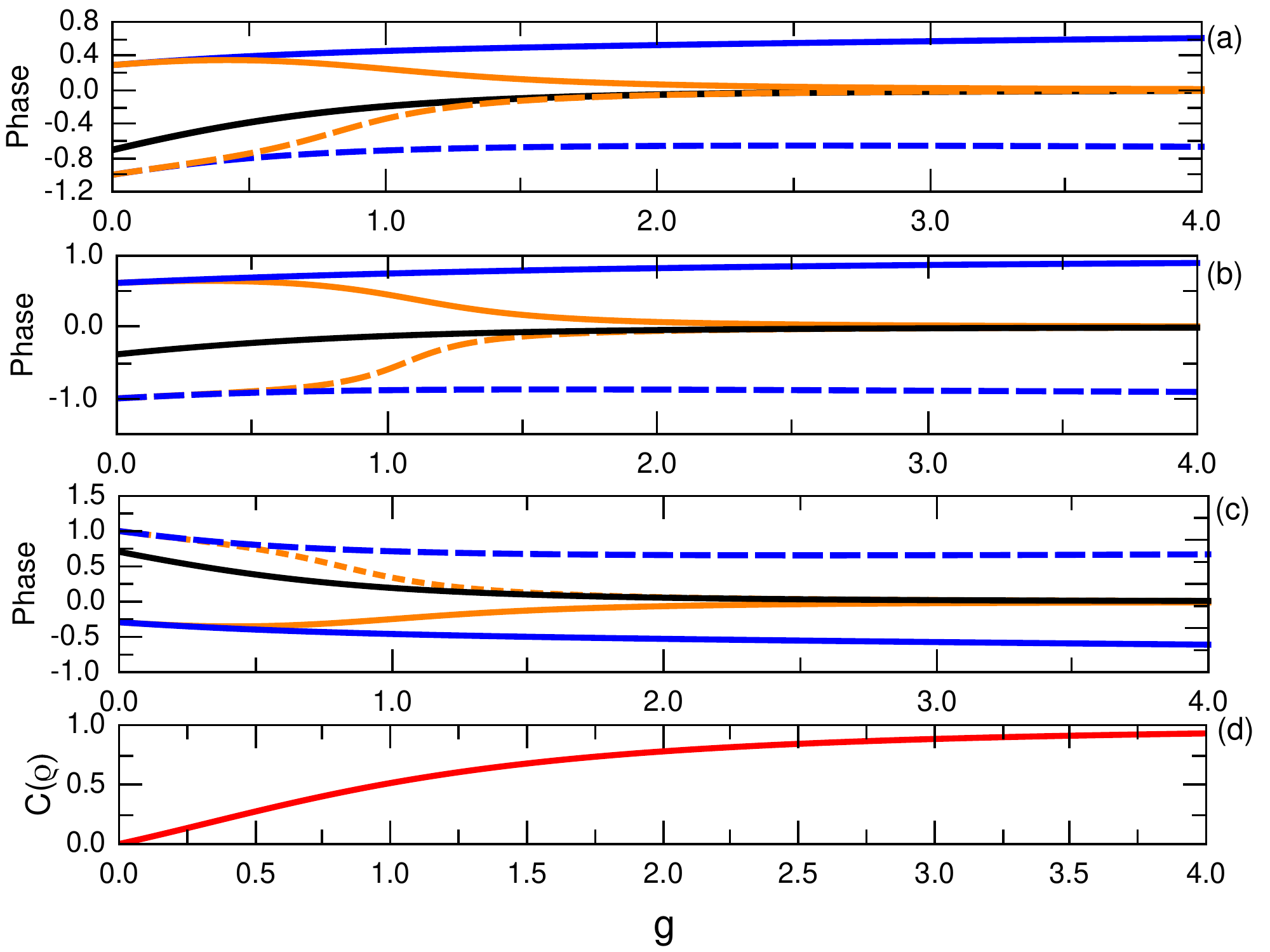}}
   \end{center}
     \caption {Comparison of the Uhlmann phase [ Eq.~(\ref{analiticalUhlmann})], $\Phi^A$ (orange solid line) and  $\Phi^B$ (orange dashed line), and the interferometric phase [Eq.~(\ref{BerrySubsystems})],  $\gamma_I^A$ (blue solid line), and  $\gamma_I^B$ (blue dashed line), as a function of $g$. Different direction where chosen: (a) $\theta=\pi/4$, (b) $\theta=3\pi/8$, (c) $\theta=3\pi/4$. In panels (a)-(c)  we also include for comparison the Berry phase of the composite system, $\gamma$ [Eq.~(\ref{Berry_uj})] (black solid line), and in (d) we present the concurrence, ${\cal C}(\rho)$ (red solid line). }
          \label{UandBColSubAyBNOthermal}
           \end{figure}

In Fig.~\ref{colormapjeq2} we argued that for large values of the concurrence (high entanglement) that $\Phi^B\rightarrow \gamma$ for all directions of the field, $\theta$. We now consider several fixed directions to further emphasize this result. 
In Fig.~\ref{UandBColSubAyBNOthermal}(a)-(c) we show the behavior of Uhlmann and interferometric phase for the subsystems, given respectively by $\Phi^s$ [Eq.~(\ref{analiticalUhlmann})] and $\gamma_I^s$ [Eq.~(\ref{BerrySubsystems})], as a function of $g$, for different direction of the field.  
We demonstrate  Fig.~\ref{UandBColSubAyBNOthermal}(a)-(c) that for small values of the coupling ($g\lesssim 0.25$), both  Uhlmann and interferometric phases coincide for each of the subsystems.
These phases coincide exactly when the system is completely uncoupled ($g=0$).
%
We also observe that for larger values of the coupling parameter, the phases differ notably from each other.  
Unlike the interferometric phases of the subsystems, which tend to be almost constant for large values of the coupling, the Uhlmann phases decrease rapidly as a function of $g$. We also emphasize on the peculiar behavior
of the Uhlmann phase  $\Phi^B$, in the large $g$ limit. We show that the latter phase tends to the Berry phase of the composite system $AB$, $\gamma$, calculated using Eq.~(\ref{Berry_uj}).
By measuring the concurrence given by Eq.~(\ref{concurrence}) associated to the \textit{ground state},  as shown in  Fig.~\ref{UandBColSubAyBNOthermal}(d), 
we can characterize the general behavior of the Uhlmann and interferometric phases of the subsystems.
We argue that (i) for \textit{weak entanglement}, the Uhlmann and interferometric phases in each subsystem coincide with each other, and that (ii) for \textit{strong entanglement}, the Uhlmann and interferometric phases in each subsystem are in general different form each other.
It is in this regime of \textit{strong entanglement} where we observe that $\Phi^B\rightarrow \gamma$.  
That is, the Berry phase of the composite system $\gamma$,  becomes accessible by measuring the Uhlmann phase of the subsystem $B$,  $\Phi^{B}$ in the strong coupling regime.
The observed behavior for the Uhlmann phase in the weak and strong entanglement regimes hold for different values of $\theta$.
Similar results are obtained for the eigenstates $j=1$ (not shown here) since the phases only differ by a minus sign.
\begin{figure}[H] 
\begin{subfigure}{4.7cm}
\includegraphics[width=4.5cm]{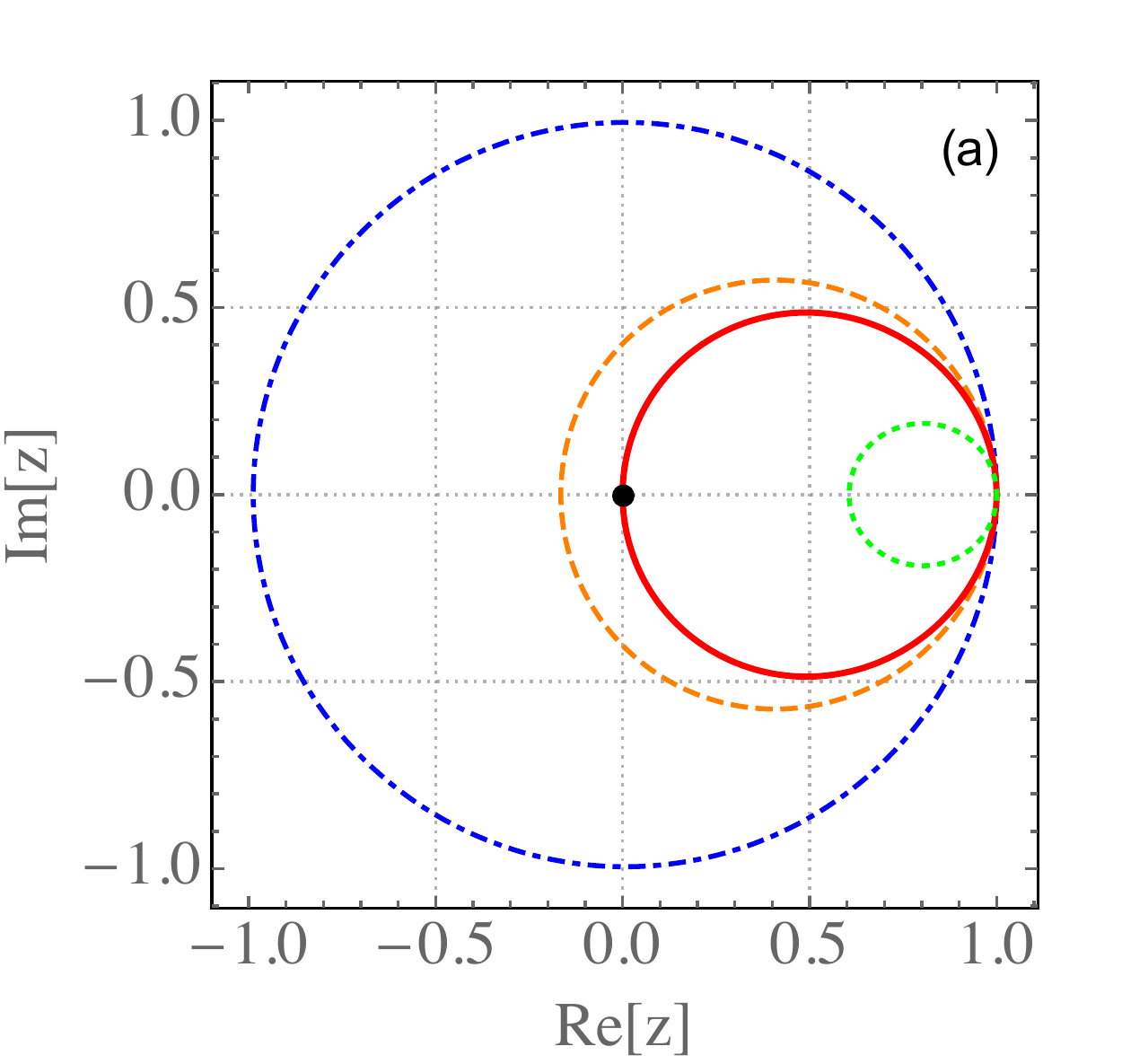}
\end{subfigure}%
\begin{subfigure}{4.7cm}
\includegraphics[width=4.5cm]{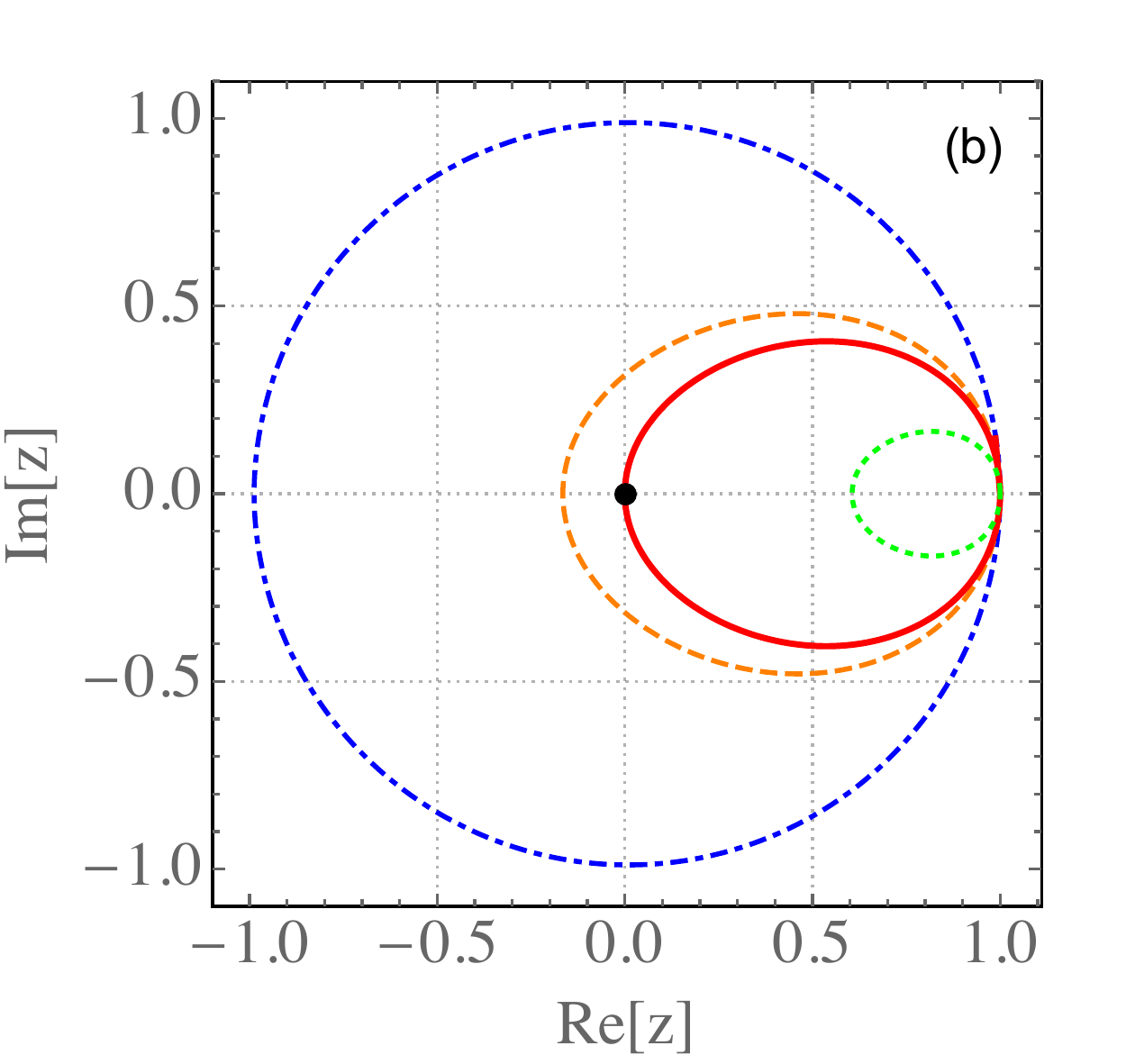}
\end{subfigure}
\vspace{10pt}
\caption[]{Curves (a) $z^A(\theta)$ and (b) $z^B(\theta)$, at several values of the coupling strength; $g=0.1$ (blue dashed dotted line), $g=1$ (orange dashed line), $g=g_c=2/\sqrt{3}$, (red solid line), and $g=2$ (green dotted line). There is a phase singularity at the root of the polynomial $-U_1(z)$, located at $z=0$ in this diagram (black dot), which is encircled by the curve $z^s(\theta)$ whenever $g<g_c$. The winding number of the curve
$-U_1(z^s(\theta))$ is 1 when $g<g_c$, and 0 otherwise.}
\label{Argand}
\end{figure}
The phase transition observed in Fig.~\ref{subsyAyBpientre2_allj}(a) can be characterized by a change of a winding number.
To see this, we follow Ref.~\onlinecite{PhysRevA.103.042221} and write the Uhlmann phase (\ref{analiticalUhlmann}) in the form
$\Phi^s={\rm Arg}\{-U_1(z^s(\theta))\}$ where  $U_1(z)=2z$ is the second-kind Chebyshev polynomial of order one, 
with argument $z^s(\theta)=\{\cos(\pi r_s)+i\, [(\bar{\gamma}^s-\pi)] \sin(\pi r_s)/\pi r_s\}/2$.
In Fig.~\ref{Argand} we  plot the curve $z^s(\theta)$ for several values of the coupling strength $g$.
We note that the only root of the function $-U_1(z^s(\theta))$, the point $z=(0,0)$, is encircled by $z^s(\theta)$
whenever $g<g_c$, and lies outside it for $g>g_c$. Thus, according to the Argument Principle of complex analysis  \cite{VisualComplexFunctions}, the curve $-U_1(z^s(\theta))$ has a winding number of 1 about zero for $g<g_c$ or 0 otherwise, which is the number of times the phase $\Phi^s$ (\ref{analiticalUhlmann}) changes from 0 to $2\pi$. 
The phase vortex at $(\theta,g)=(\pi/2,g_c)$ reveals the threshold for a transition between topological orders with different winding numbers, in a very similar manner to the appearance of critical temperatures for Uhlmann phase transitions of a spin-$j$ particle in an external magnetic field \cite{PhysRevA.103.042221}.

\subsection{Depolarizing channel with $q\neq 0$}\label{sec:subsystemsAandBqneq0}

We investigate the geometric phases of the subsystems corresponding to the \textit{depolarizing channel} (\ref{werenerdensity}) with a depolarization strength, $q\neq 0$. 
We follow the procedure of Sec.~\ref{subsec:depq0} 
to obtain the density matrices for the subsystems $A$ and $B$ by computing the trace of $\rho_{d}$ over $A$ (or $B$), to obtain $\rho^{s}_{d}$, which are related to $\rho^{s}$ as follows: $\rho^{s}_{d}=(q/2)\,\mathds{1}_2+(1-q)\,\rho^s$.
The density matrix, $\rho^{s}_{d}$, has the same  explicit form as (\ref{bastardnotation}), with new coefficients, $a_{d,s}$, and $c_{d,s}$, which are related to the coefficients of $\rho^{s}$, \textit{i.e.}  $a_{d,s}=(q/2)+(1-q)a_{s}$, and $c_{d,s}=(1-q)c_{s}$. 
Also, since the Hermitian operators $\rho^{s}$ and $\rho^{s}_{d}$ share the same  eigenfunctions (\ref{eigenvalspin}), the corresponding eigenvalues are $p_{d,s,l}=(q/2) + (1-q) p_{s,l}$,
or equivalently, they can be obtained from (\ref{eigenvalspin}) by making the substitutions  $a_{s}\rightarrow a_{d,s}$, and $c_{s}\rightarrow c_{d,s}$.
The eigenvalues  $p_{d,s,l}$ also satisfy the same properties as  $p_{s,l}$. 
Thus, the Uhlmann phase for a depolarizing channel,  $\Phi_{d}^{s}$ of the subsystem $s$, associated to the $j$-th eigenstate can be calculated  analytically by following the procedure of Sec.~\ref{subsec:depq0}. With the help of the identity  ${\rm det}[\rho^{s}_{d}]=(q/2)(1-q/2)+(1-q)^2\,{\rm det}[\rho^{s}]$, which relates the determinants of the  depolarizing and the pure state matrices, we obtain, 
\begin{equation}
\Phi_{d}^{s}={\rm Arg}\left\{-\cos(\pi r_{d}^s)-i\,(1-q)\, \left[\bar{\gamma}^s-\pi\right]\frac{\sin(\pi r_{d}^s)}{\pi\, r_{d}^s} \right\},  \\
\label{analiticalUhlmannW}
\end{equation}
where
\begin{equation}
r_{d}^s= \left(1-\,\gamma^{s,1} \gamma^{s,2}\,[1-q]^2\left[{1-\cal C}^2(\rho)\right]/\pi^2 \right)^{1/2}.
\label{rsdepolarizing}
\end{equation}
Note that  Eq.~(\ref{rsdepolarizing}) depends on the degree of mixing in the  Bloch representation of the density matrix from the depolarizing channel,  $\rho^{s}_{d}=[\mathds{1}_2+R_d\,\hat{\bm{n}}_s\cdot\bm{\sigma}]/2$, where the effect of $q$ is to reduce the radius of the Bloch sphere (depolarizing effect) as $R_d=(1-q)R=(1-q)[1-{\cal C}^2(\rho)]$. 
Note also that in the case $q\rightarrow 0$, we recover the results for the Uhlmann phase for the \textit{pure state} given by Eqs.~(\ref{analiticalUhlmann}) and (\ref{lars}). 

\begin{figure}[H]
 \begin{center}
  {\includegraphics[angle=0,width=3.5in]{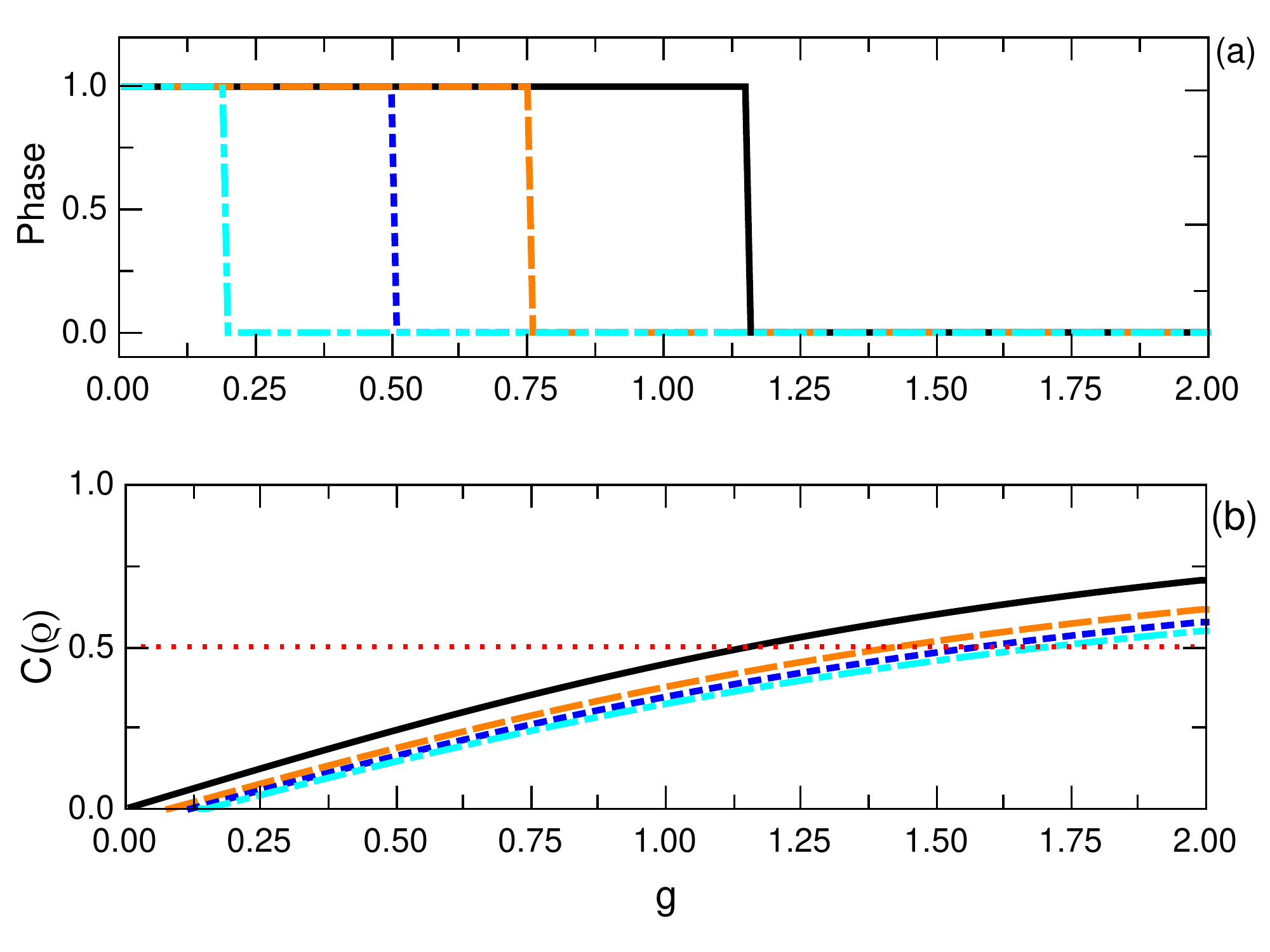}}
    \end{center}
     \caption {(a) Comparison of the Uhlmann phases $\Phi^s_{d}$  [Eq.~(\ref{UhlmansAyBpientre2W})], as a function of the coupling parameter $g$, for different values of the depolarizing strength, $q$: 0.0 (black solid line), 0.075 (orange dashed line), 0.1073 (blue dotted line), and 0.13 (cyan dashed dotted line). The topological phase transition occurs at a critical value $g_{d,c}$, given by Eq.~(\ref{gwc}). (b) Concurrence, ${\cal C}(\rho_{d})$, for the $j$-th eigenstate of depolarizing channel, for the same values of $q$. The direction $\theta=\pi/2$ was chosen for this calculation. }
          \label{subsyAyBpientre2_alljWerne}
           \end{figure}

The Uhlmann phase in  Eq.~(\ref{analiticalUhlmannW}) can be simplified along the direction $\theta=\pi/2$, because  $a_{d,s}=1/2$,  $\delta_s=0$, and $\gamma^{s,l}=\bar{\gamma}^s=\pi$, which leads to  
\begin{equation}
\Phi_{d}^{s}={\rm Arg}\{-\cos\left[\pi\,r_{d}^s\right]\},    
\label{UhlmansAyBpientre2W}    
\end{equation}
and also the factor in (\ref{rsdepolarizing}) is simplified to 
$r^d_s=[1-(1-q)\,R_d]^{1/2}$. 
We observe that the topological transition occurs at $r_s^d=1/2$, leading us to the condition,
\begin{equation}
(1-q)^2\,[1-{\cal C}^2(\rho)]=3/4.
\label{conditionrsd}    
\end{equation}
The latter expression, with the help of Eq.~(\ref{Cderho}), allows us to calculate the critical value of the coupling as a function of the depolarizing parameter, $q$,
\begin{equation}
g_{d,c}=g_c\,\sqrt{4q^2-8q+1},
\label{gwc}
\end{equation}
showing that the topological transition of the Uhlmann phase can be controlled by tuning the parameter $q$. 
By inspecting (\ref{gwc}), we find that the maximum 
of the critical coupling, $g^d_c=g_c$, occurs at $q=0$, and its minimum value $g^d_c=0$ is reached when $q=1-g_c^{-1}$.
Therefore, the phase transition is restricted to the following range values: $q\in(0,1-g_c^{-1}]$. 
In Fig.~\ref{subsyAyBpientre2_alljWerne}(a) the Uhlmann phase $\Phi^s_{d}$  [Eq.~(\ref{UhlmansAyBpientre2W})] is plotted for different values of $q$.
By controlling the polarization strength,  we can shift the Uhlmann phase transition of the system along the interval $g_{d,c}\in[0,g_c]$. 

Note that  $\Phi_{d}^{s}$ depends explicitly  on the concurrence of the \textit{pure state}, $C(\rho)$. 
We now investigate the degree of entanglement in the composite system $AB$, which is required to observe the phase transitions in the corresponding subsystems.  
We proceed by showing the following relationship  between the concurrences of the mixed and pure states,
\begin{equation}
C(\rho_d)=\mbox{max}\left\{0,(1-q)\,C(\rho)-q/2 \right\},
\label{concurrenceanzat}    
\end{equation}
valid for any value of $q$, $g$, and $\theta$. 
We now use  (\ref{conditionrsd}) and  (\ref{concurrenceanzat}) to find the concurrence for the phase transition as a function of $q$,
\begin{equation}
C(\rho_d)=\mbox{max}\left\{0,\left[\left(g_{d,c}/g_c\right)-q \right]/2 \right\}.
\label{concurrenceanzatzbis}    
\end{equation}
Since this transition is restricted to a depolarization strength 
$0 < q \leq 1-g_c^{-1}$, these values allow to find from (\ref{concurrenceanzatzbis}) the upper and lower bounds of the concurrence: $0< C(\rho_d) \leq 1/2$.
This is shown in Fig.~\ref{subsyAyBpientre2_alljWerne}(b), where we plot $C(\rho_d)$ for different values of $q$ for $\theta=\pi/2$.

\begin{figure}[H] 
\begin{subfigure}{4.7cm}
\includegraphics[width=4.5cm]{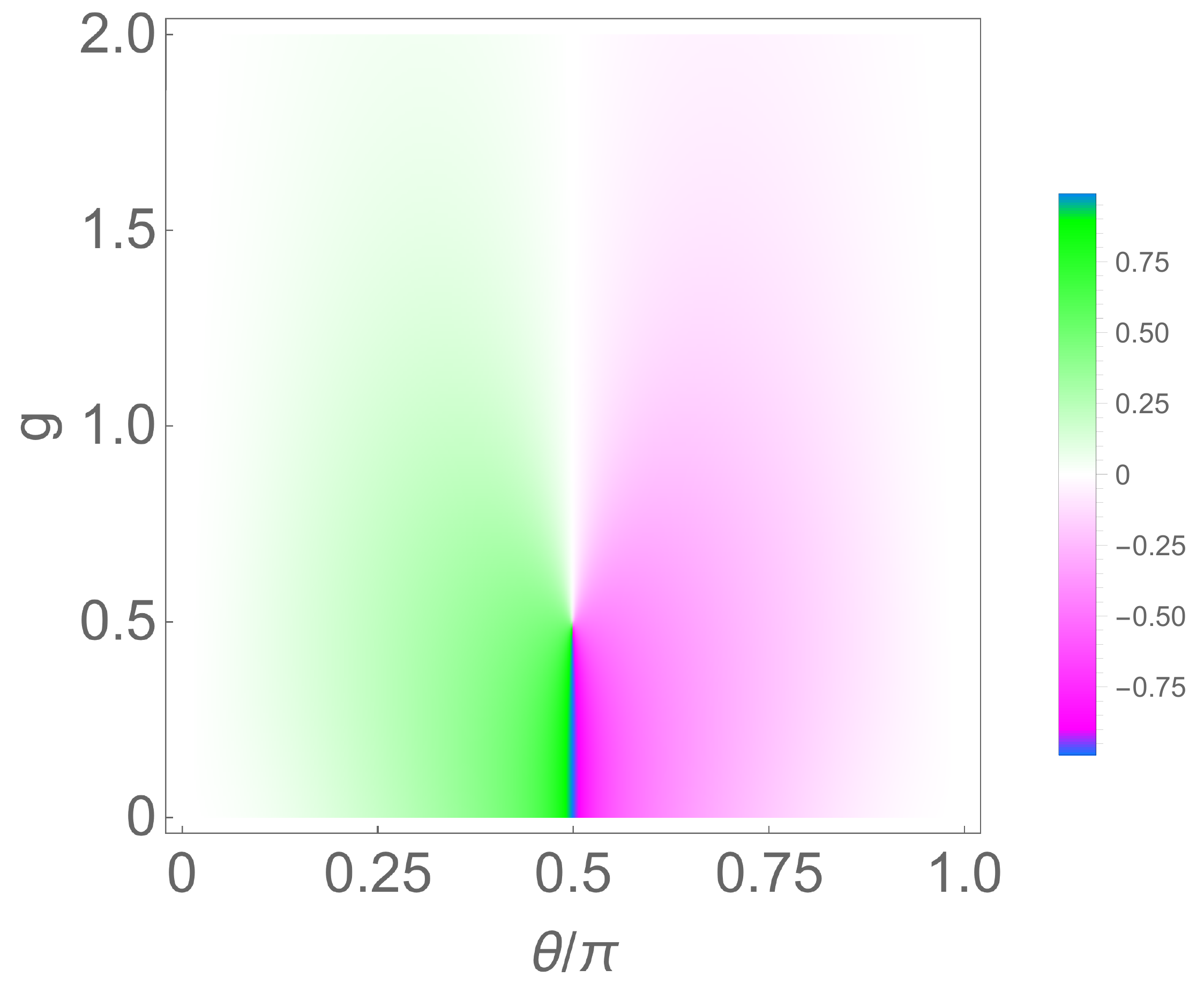}
\caption{Uhlmann phase $\Phi_{d}^A$. }
\end{subfigure}%
\begin{subfigure}{4.7cm}
\includegraphics[width=4.5cm]{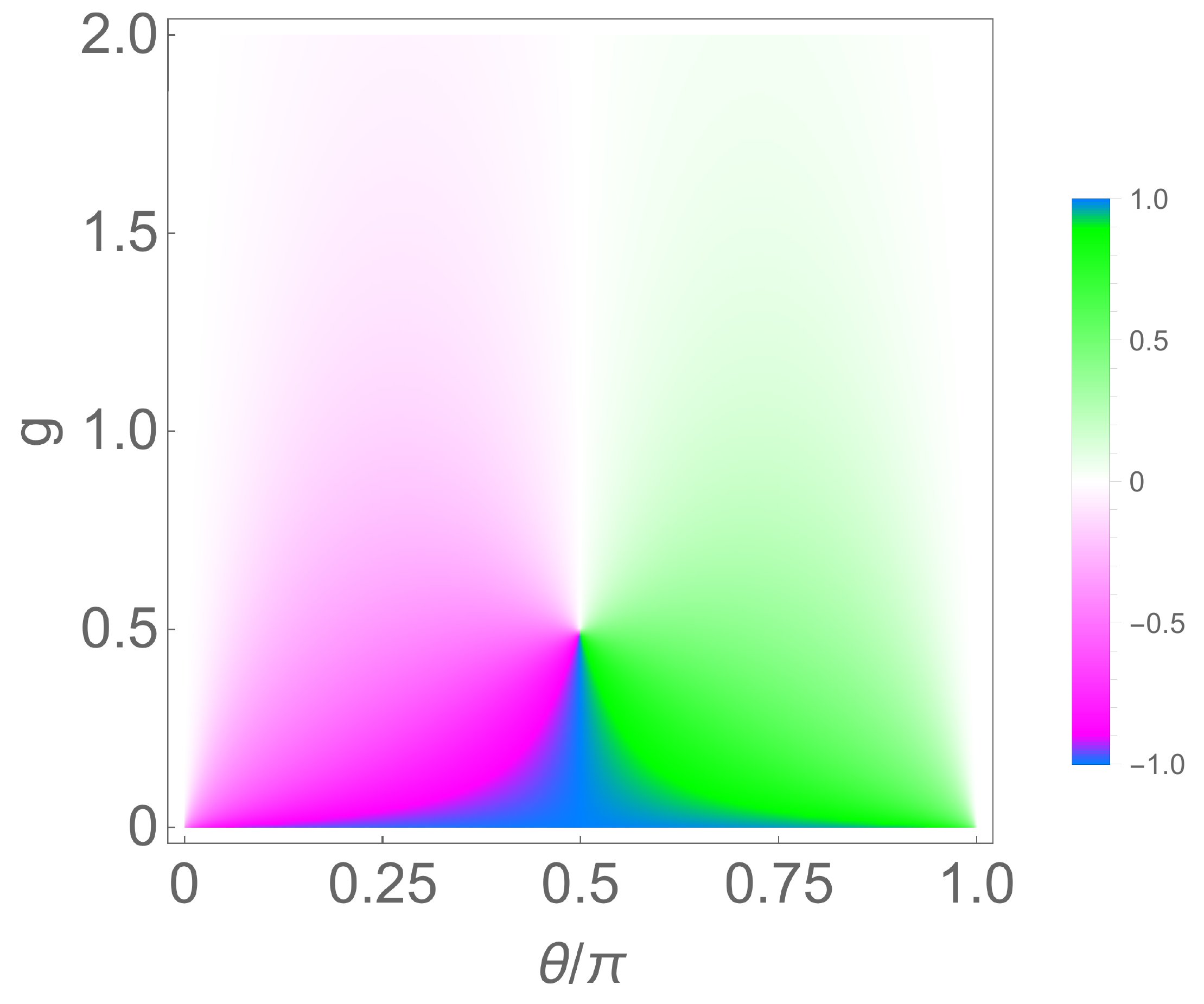}
\caption{Uhlmann phase $\Phi_{d}^B$.}
\end{subfigure}\vspace{10pt}
\begin{subfigure}{4.7cm}
\includegraphics[width=4.5cm]{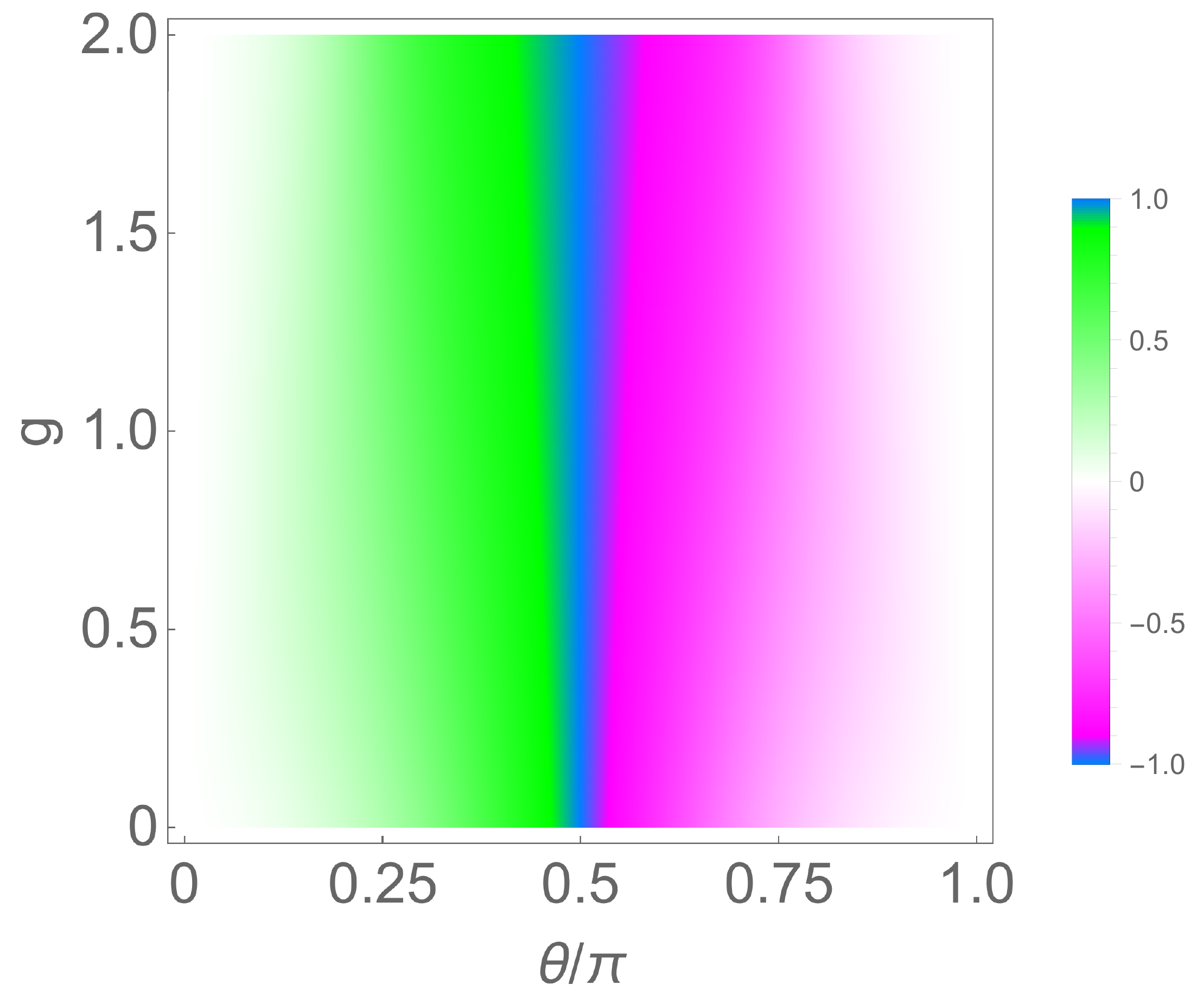}
\caption{Interferometric phase $\gamma_{I,d}^A$. }
\end{subfigure}%
\begin{subfigure}{4.7cm}
\includegraphics[width=4.5cm]{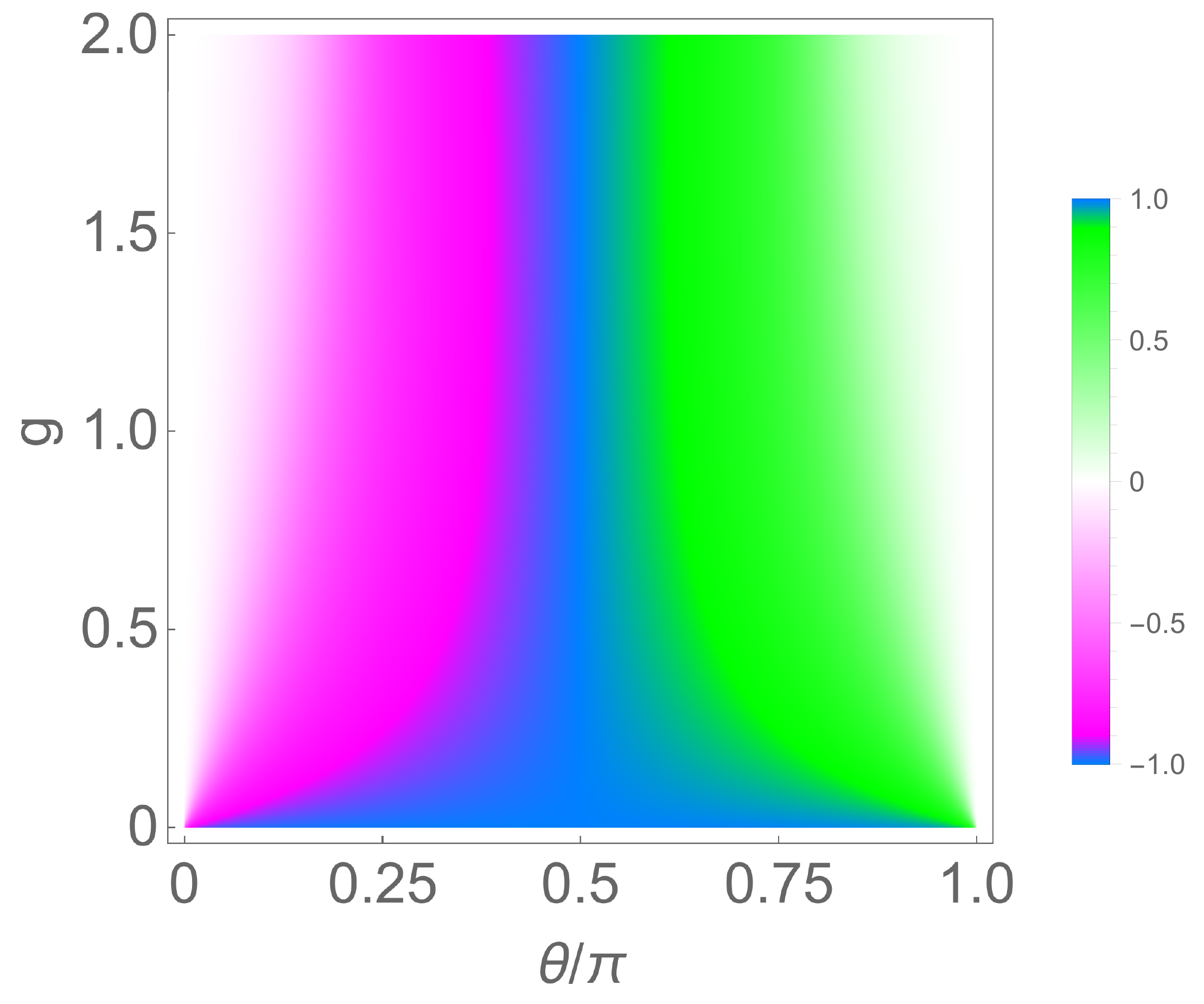}
\caption{Interferometric phase  $\gamma_{I,d}^B$.}
\end{subfigure}\vspace{10pt}
\begin{subfigure}{4.7cm}
\includegraphics[width=4.5cm]{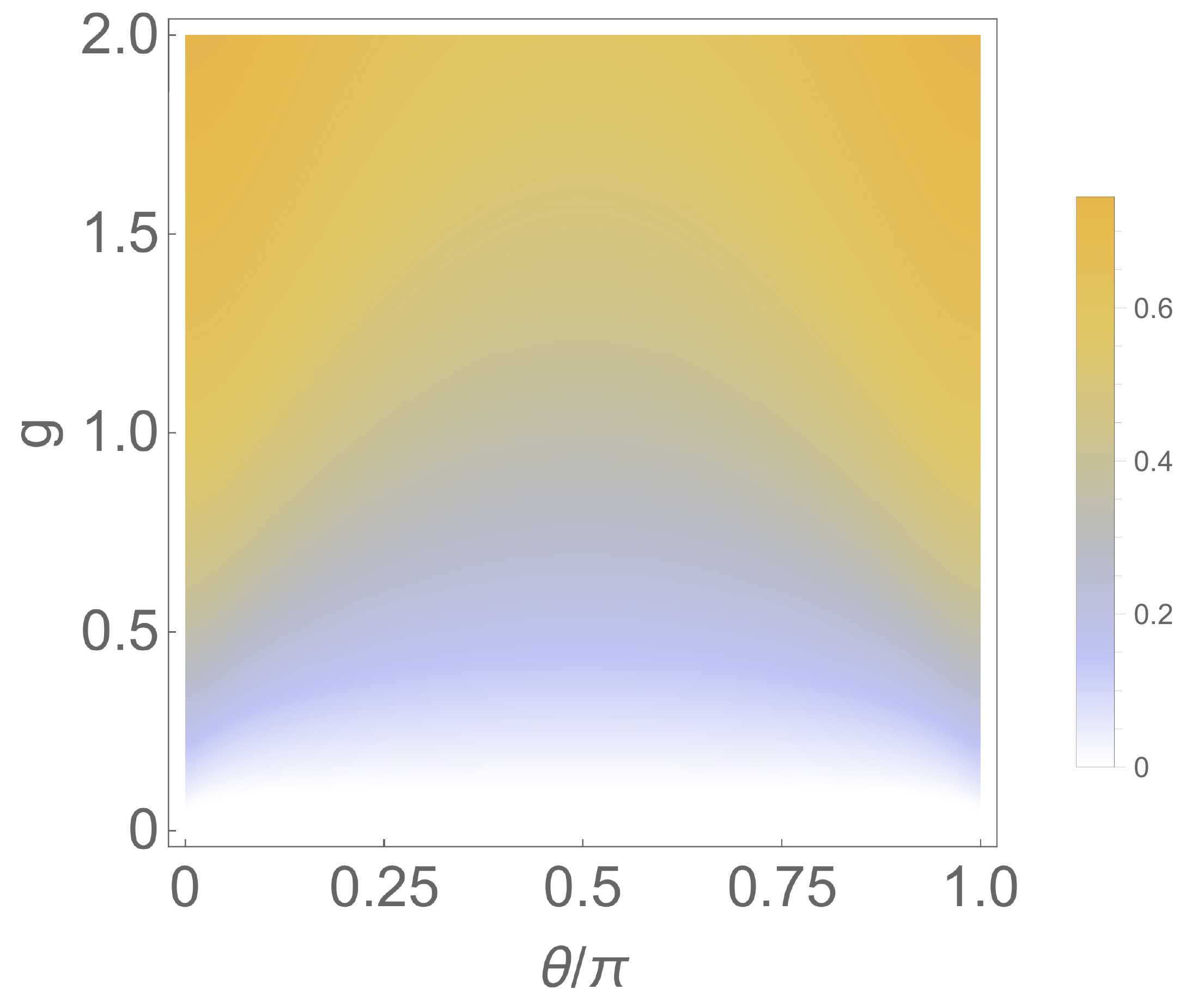}
\caption{Concurrence ${\cal C}(\rho_d$). }
\end{subfigure}
\caption[]{Color density maps of (a)-(b) the Uhlmann phase $\Phi^s_{d}$ [Eq.~(\ref{analiticalUhlmannW})] and (b)-(c) interferometric phase $\gamma_{I,d}^s$ [Eq.~(\ref{BerrySubsystems_dep})], for a depolarizing channel with $q=0.1073$ (yields $g_c^d=1/2$) as a function of  $g$, and $\theta$. We have also included  (e) the concurrence of the composite system, ${\cal C}(\rho_d)$. For this particular case the topological transition at $\theta=\pi/2$ occurs at ${\cal C}(\rho_d)=0.1628$ as computed from Eq.~(\ref{concurrenceanzatzbis}).  }
\label{colormap1AWerner}
\end{figure}

Also, as discussed in Sec.~\ref{subsec:depq0}, we expect the appearance of a vortex in the density maps of the Uhlmann phase of the subsystems along the direction $\theta=\pi/2$, which rapidly vanishes as the parameter $q$ is increased.  
In Figs.~\ref{colormap1AWerner}(a)-(b),  we show the behavior of the Uhlmann phase $\Phi^s_{d}$ [Eq.~(\ref{analiticalUhlmannW})] for a particular value of $q\neq 0$, to emphasize the change in position of the vortex. 
The vortex disappears when the value of  $q=(1-g_c^{-1})$ is reached (not shown).
In Figs.~\ref{colormap1AWerner}(c)-(d) we have also included for comparison the interferometric phase of the subsystems, 
\begin{equation}
\gamma_{I,d}^s={\rm Arg}\left\{\sum_{l=1}^{2} p_{d,s,l}\,e^{i\,\gamma^{s,l}}\right\}.
\label{BerrySubsystems_dep}    
\end{equation}

Our results for the Uhlmann phase show that it
has a structure that exhibits topological phase transitions between orders of different winding numbers.
These transitions can alternatively be characterized by system concurrence. It is important to note that for this particular model of entangled fermions, although the Berry phase for mixtures given by the interferometric phase \cite{PhysRevLett.85.2845} coincides with the Uhlmann phase in some regimes of interest, it does not exhibit a phase vortex profile.

\section{Conclusions}\label{sec:conclusions}

We investigate the topological Uhlmann phase acquired by the composite system $AB$ and its corresponding  subsystems for  two interacting fermions with spin-$\frac 1 2$, where one of the spins is driven by a time-dependent magnetic field, with a coupling parameter, $g$. 
We use a depolarizing channel model to explore  the Uhlmann phase $\Phi^s_{d}$, corresponding to subsystems $s=A,B$, for entangled mixed states in our two-qubit setup, and derive exact analytical solutions for  $\Phi^s_{d}$ [Eq.~(\ref{analiticalUhlmannW})]. 
We find that the Uhlmann phase depends explicitly on the depolarization strength, $q$,
%
%
and on different combinations of phases.
%
%
%
%
%
Moreover, it is shown that $\Phi^s_{d}$ exhibits an explicit connection to the concurrence, ${\cal C}(\rho_d)$, which provides a measure of the degree of entanglement in the composite system.
Our results show that in general,  $\Phi^s_{d}$ exhibits a topological structure, with a phase vortex profile associated to a phase transition along the direction $\theta=\pi/2$.
We compare our results for the Uhlmann phase with those obtained by  alternative formulations for a geometric phase involving mixtures, such as the interferometric phase \cite{PhysRevLett.85.2845}, and find some important differences. 
For the \textit{pure state} case ($q=0$) involving the ground state of the system, we find that $\Phi^s$ [Eq.~(\ref{analiticalUhlmann})], and $\gamma_I^s$ [Eq.~(\ref{BerrySubsystems})], agree with each other in the low-coupling regime (weak entanglement). This is not the case for the large-coupling regime (strong entanglement), where both geometric phases differ noticeably from each other. 

Regarding the Uhlmann phase transitions along $\theta=\pi/2$, we demonstrate that (i) for the \textit{pure state} case ($q=0$) such a transition occurs at a fixed critical coupling, $g_c=2/\sqrt{3}$. (ii) For the \textit{depolarizing channel} case ($q \neq 0$), we show that a whole range of transitions at critical values $g_c^d$ 
are allowed, as long as the critical values of the depolarization strength are restricted to $q\in(0,1-g_c^{-1}]$.
We derive a simple formula to determine the corresponding critical coupling, $g_c^d$ [Eq.~(\ref{gwc})], as a function of $q$. 
It is also demonstrated that each of these critical values of the depolarization, yield a concurrence of the composite system that is bounded as $0< {\cal C}(\rho_d) \leq 1/2$. 
In other words, the Uhlmann phase transitions of the subsystems, and the corresponding concurrence of the composite system, can be tuned by controlling the depolarization strength, $q$.

We hope that our results may stimulate further studies of the Uhlmann phase in composite entangled systems with temperature-induced topological transitions.

\section{Acknowledgements}\label{acknowledgements}

DMG acknowledges support from CONACYT (M\'exico). JV thanks CNyN--UNAM for their kind hospitality during the sabbatical leave, and also I. Maldonado for fruitful discussions.

\end{document}